 \documentclass[12pt,a4paper]{article}
 \pdfoutput=1
 \usepackage{jheppub}

\usepackage{amsmath}
\usepackage{bm}
\usepackage{wrapfig}

\newcommand{\be}{ \begin{equation}}
\newcommand{\ee}{\end{equation}}
\newcommand{\bea}[1]{\begin{eqnarray}\label{#1} }
\newcommand{\eea}{\end{eqnarray}}

\def\ZZZ{{\hskip-3pt\hbox{ Z\kern-1.6mm Z}}}
\def\zzz{{\hskip-3pt\hbox{ z\kern-1mm z}}}

\newcommand{\half}{{1\over 2}}

\def\one{{\hbox{ 1\kern-.8mm l}}}
\def\zero{{\hbox{ 0\kern-1.5mm 0}}}

\definecolor{rust}{rgb}{0.8,0.2,0.2}
\definecolor{green}{rgb}{0.1,0.8,0.2}

\def\hpeak{h_\text{peak}}
\def\hmax{h_\text{max}}
\def\barh{{\bar h}}
\def\AdS#1{AdS$_{#1}$}

\def\fig#1{Fig.\,\ref{#1}}

\title{The Spectrum of Light States in Large N Minimal Models}

\author{Matthias R.\ Gaberdiel$^a$,}
\author{ Rajesh Gopakumar$^b$,}
\author{  Mukund Rangamani$^c$}

\affiliation[a]{Institut f\"ur Theoretische Physik, ETH Zurich, \\
$\;$ CH-8093 Z\"urich, Switzerland }
\affiliation[b]{Harish-Chandra Research Institute, \\ Chhatnag Road, Jhusi,
Allahabad, India 211019}
\affiliation[c]{ Centre for Particle Theory \& Department of Mathematical Sciences,\\
Science Laboratories, South Road, Durham DH1 3LE, UK}

\emailAdd{gaberdiel@itp.phys.ethz.ch}
\emailAdd{gopakumr@hri.res.in}
\emailAdd{mukund.rangamani@durham.ac.uk}

\abstract{$W_{N,k}$ minimal models possess an interesting class of `light' primaries which control much of the low energy density of states in the large $N$ 't Hooft limit. In this paper we conduct a detailed exploration of their distribution using a combination of numerical and analytical techniques. We also make some observations about the density of states of the full CFT. Our results appear to support the contention that there is no finite temperature analogue of the Hawking-Page transition in these systems.  } 

\keywords{}

\begin{document}
\begin{flushright} \small{DCPT-13/27} \end{flushright}

\maketitle

\flushbottom

\section{Introduction and Summary}

The holographic AdS/CFT correspondence provides a unique perspective on the dynamics of quantum fields (both with and without dynamical gravity). In order to understand how this miraculous duality can be derived from first principles one would like to have access to simple examples of quantum field theories which naturally fall within the remit of this correspondence. In recent years a class of such theories has emerged: these are the `vector-like models' with higher spin symmetries in $d=2,3$ spacetime dimensions, which have been conjectured to be dual, in an appropriate leading large $N$ limit, to a Vasiliev higher spin theory in \AdS{3} and \AdS{4} respectively. For recent reviews on higher spin holography and an exhaustive list of references, see \cite{Gaberdiel:2012uj, Giombi:2012ms, Ammon:2012wc}.

The two dimensional theories encompass a class of interacting quantum field theories which will be the focus of our current investigation: these are the $W_{N,k}$ coset minimal models. The minimal model holography relates the dynamics of these 
$W_{N,k}$ models to the dynamics of a higher spin gauge theory in \AdS{3} \cite{Gaberdiel:2010pz}. More precisely, the field theory in the 't Hooft limit $N,k \to \infty$ with $\lambda = \frac{N}{N+k}$ fixed, is dual to the Vasiliev theory with $\text{hs}[\lambda]$ gauge group coupled to a single complex scalar field of mass $m^2 = -1 + \lambda^2$
\cite{Prokushkin:1998bq,Prokushkin:1998vn}. 
Evidence for this statement comes from a delicate matching of symmetries \cite{Gaberdiel:2012ku}, comparison of the partition function in the strict planar ($N \to\infty$) limit \cite{Gaberdiel:2011zw} including a match of the asymptotic growth of states with black hole entropy \cite{Kraus:2011ds, Gaberdiel:2012yb}, agreement of three and higher point correlation functions \cite{Chang:2011mz,Ahn:2011by, Ammon:2011ua, Chang:2011vka, Hijano:2013fja} as well as a recent (bulk) one loop test \cite{Giombi:2013fka}. 

Despite these successes there are aspects of the minimal models which are as yet poorly understood in the 't Hooft limit. While we know that the spectrum of the CFT contains a higher spin current $W^{(s)}$ for each integer $s \geq 2$, a systematic understanding of the full spectrum of operators has yet to be achieved. To be sure, for finite $N$  and $k$ the spectrum of primary operators is known. It can be described in terms of representations of $\mathfrak{su}(N)_l$ for $l = k$ and $l=k+1$, respectively. One moreover has an explicit expression for the conformal dimension of the operators in terms of the group theoretic data. The main issue is to figure out how the finite number of primaries that are present for finite values of $N$ and $k$ behave in the large $N$ 't Hooft limit. 
This is somewhat challenging for two reasons: (i) the number of primaries grows quite rapidly as an exponential in $N$ in this limit (as we explain later); and (ii) certain operators acquire anomalously low conformal dimensions in the limit \cite{Gaberdiel:2010pz, Gaberdiel:2011zw, Papadodimas:2011pf} --- these have come to be known as the {\em light states}. Heuristically one can visualise the spectrum as forming a near-continuum above the vacuum. The absence of a gap in the spectrum of primaries poses novel challenges for understanding the theory and the holographic map.\footnote{We should point out that there exists a `semi-classical' limit where $N$ is kept fixed and $c\rightarrow \infty$ \cite{Castro:2011iw} in which this continuum disappears and there is a compelling match \cite{Perlmutter:2012ds, Campoleoni:2013iha} of all the primaries in the CFT with conical defect like geometries \cite{Castro:2011iw, Campoleoni:2013lma} together with perturbative scalar quanta. However, the CFT is not unitary in this limit and hence the exact nature of the holographic map is unclear. We will not be considering the semi-classical limit in this paper though it would be also interesting to study the density of states in that limit. See \cite{Chang:2013izp} for a different bulk interpretation of primaries in the 'tHooft limit based on the classification in terms of single particle and multi particle states \cite{Jevicki:2013kma, Chang:2013izp}.}

One natural question which arises in the holographic context is whether these theories exhibit a phase transition in the canonical ensemble. From a field theory perspective, this would correspond to the change in the nature of the primaries (hence the representations of $\mathfrak{su}(N)_l$) which provide the dominant saddle point to the canonical partition function. In order for this behaviour to be under control, one requires the transition temperature to be ${\cal O}(1)$ in the 't Hooft limit. This is, of course, what we see in conventional examples of AdS/CFT without higher spin fields; for example the Hawking-Page transition occurs at $T\sim {\cal O}(1)$ in the planar limit. This naively appears not to be the case in the higher spin theories: in the case of vector models in $d=3$, it 
was argued in \cite{Shenker:2011zf} that the phase transition temperature scales with a positive power of $N$,
namely  $T \sim N^\frac{1}{2}$. A similar analysis for the coset minimal models was undertaken earlier in the limit $\lambda \to 0 $ in \cite{Banerjee:2012aj}, with the conclusion that there is no transition at temperatures of ${\cal O}(1)$. The presence of the light states appears to smooth out the possibility of a phase transition. Indeed, the continuum of low-lying states seems to be responsible
for not having the abrupt jump coming from the discretuum of black hole micro states (above the black hole threshold) that 
other theories of gravity on AdS$_3$ exhibit. This is what we would like to closely scrutinise in this paper. 

To get a more precise picture of the spectrum of these coset CFTs we undertake an exploration of the primary operators in the 
$W_{N,k}$ theory employing a variety of methods. Firstly, the basic problem of the operator spectrum can be phrased as a counting problem, i.e., we can enumerate the number of primaries ${\cal D}(h)$ with a given conformal dimension $h$. Since $h$ is determined in terms of some group theoretic data, this particular question is well suited to numerical exploration for finite $N$ and $k$. By explicitly writing down all the allowed representations and computing the conformal dimensions, we are in a position to obtain some intuition for ${\cal D}(h)$. While this exercise can be done for the full spectrum, it turns out to be much simpler and more effective in constraining the spectrum of the light states. Part of the reason is that there are (exponentially) fewer light states and one can therefore numerically test this part of the spectrum for higher values of $N$ and $k$. But more crucially, as we explain, the spectrum of light states can be mapped (for any value of $\lambda$) to a much simpler  problem, that of free fermions in one dimension.

The data contained in ${\cal D}(h)$ can, by a Legendre transformation, be packaged into a canonical partition function $Z(\beta)$ with $\beta = T^{-1}$.  The group theoretic data for a light state can be distilled down to a single $\mathfrak{su}(N)$ Young tableaux. There is of course an intimate connection between the row labels of a Young tableaux and free fermions which arises due to the canonical ordering on the rows. This has in fact been seen in many earlier studies of AdS/CFT, e.g., in \cite{Berenstein:2004kk, Lin:2004nb}. Here, we encounter however a novel variant of the conventional problem since our fermions have their momenta constrained. Aided by the numerical exploration and the mapping to the free fermion problem, we chart out the spectrum of the light states in some detail. 

We should mention though that in order to make a definitive statement about a phase transition, 
we need to be able to control also the {\it non-light states}, as well as  the contribution of the 
descendants to the partition function. These two turn out to be quite hard to pin down. While the numerical investigation does indeed give some intuition for the non-light states, understanding the contribution of the descendants turns out to be more involved. We give some preliminary estimates for both of these, which support the thesis that these theories do not undergo a phase transition at $T \sim {\cal O}(1)$.

The outline of the paper is as follows: we begin in \S\ref{sec:gen} with a review of the general features of the $W_{N,k}$ minimal model spectrum that will play an important role in our analysis. We then go on in \S\ref{sec:numerics} to describe the results of numerical experiments of the light state part of the spectrum. After providing a heuristic explanation for some of the features of the light states in \S\ref{sec:lam01}, we turn to the task of mapping the light states to the free fermion problem in \S\ref{sec:ff}. In particular, we analyse there
the resulting statistical mechanical system and argue for the absence of any sign of non-analytic behaviour indicative of a phase transition. In \S\ref{sec:restspec} we turn to the rest of the spectrum, giving salient features of the non-light primary spectrum and some basic results for the descendant contribution. We conclude with some open questions and  further thoughts in \S\ref{sec:discuss}. Some technical details of the free fermion problem are collected in Appendix \ref{sec:lowT}.

\section{Generalities}\label{sec:gen}

Let us consider the $W_{N,k}$ coset models defined by 
\begin{equation}\label{coset}
\frac{\mathfrak{su}(N)_{k} \oplus \mathfrak{su}(N)_1}{\mathfrak{su}(N)_{k+1}} \ ,
\end{equation}	
whose primary operators are labelled by $(\Lambda^+;\Lambda^-)$ with
$\Lambda^\pm$ being integrable highest weight representations of 
$\mathfrak{su}(N)_{k}$ and $\mathfrak{su}(N)_{k+1}$, respectively.\footnote{Given these two representations, 
there exists always a unique representation of $\mathfrak{su}(N)_1$ which satisfies the coset selection rule.}  
The labeling in terms of the pairs $(\Lambda^+;\Lambda^-)$ is $N$-fold redundant since we have the field identifications
\begin{equation}\label{fieldid}
(\Lambda^+;\Lambda^-) \cong (J\Lambda^+; J\Lambda^-) \ , \quad J\in \mathbb{Z}_N \ ,
\end{equation}
where $J$ is the automorphism
\begin{equation}
\label{Jaction}
J:\; \left[\Lambda_0; \Lambda_1, \ldots, \Lambda_{N-1} \right] \;\;\mapsto\;\; 
\left[\Lambda_1; \Lambda_2, \ldots,\Lambda_{N-1}, \Lambda_0\right] \ .
\end{equation}	
Here $\Lambda_i$ are the Dynkin labels of $\Lambda$, with 
$\Lambda_0$ being the affine Dynkin label. 

Apart from their presence in the automorphism action, the affine Dynkin labels $\Lambda^\pm_0$ can be ignored for the most part. 
The representations $\Lambda^\pm$ can then be viewed as corresponding to representations of the finite dimensional Lie algebra $\mathfrak{su}(N)$ --- we will henceforth use these symbols to denote these finite dimensional representations, without hopefully causing confusion. However, not all representations of $\mathfrak{su}(N)$ define
integrable highest weight representations of $\mathfrak{su}(N)_k$ and $\mathfrak{su}(N)_{k+1}$, respectively; the relevant 
conditions are 
\begin{equation}
\sum_{i=1}^{N-1} \, \Lambda^+_i \leq k \ , \qquad \sum_{i=1}^{N-1} \, \Lambda^-_i \leq k +1 \ .
\label{lampmk}
\end{equation}	
We will often map the Dynkin labels to a Young diagram and note that the first constraint above (for $\Lambda^+$)
restricts the length of the first row of the diagram
to be no longer than $k$. Thus the representations we are interested in correspond to those Young diagrams
that fit into a $k \times (N-1)$ rectangle for $\Lambda^+$, and similarly for $\Lambda^-$. 

The conformal weight of the primary corresponding to $(\Lambda^+;\Lambda^-)$ equals 
\begin{equation}
h(\Lambda^+; \Lambda^-) = \frac{1}{2 \, p\, (p+1)} \left[\bigg| (p+1) \left(\Lambda^+ + \rho \right)  
- p\left(\Lambda^- + \rho \right) \bigg|^2  - \rho^2 \right] \ ,
\label{hdef}
\end{equation}	
where $p=k+N$, and $\rho$ is the Weyl vector of $\mathfrak{su}(N)$. We are eventually 
interested in taking the 't~Hooft limit defined as 
\begin{equation}
k \to \infty \ , \quad N \to \infty \ , \qquad \lambda = \frac{N}{N+k} = \text{fixed} \ . 
\end{equation}	
In the 't Hooft limit there are two distinct classes of primaries: (i) {\em light states} and (ii) {\em non-light states}. The former which we focus on for much of our discussion are named so because some of them have anomalously low conformal dimensions $\propto \frac{1}{N}$. 

\subsection{The Light States}

The light  states are characterised by the property that
they have a representative (after a suitable field identification (\ref{fieldid})) with $\Lambda^+=\Lambda^-$.\footnote{Note that
there is always at most one representative that has this property.}  
As a result, in the 't~Hooft limit their conformal dimension equals 
\begin{equation}\label{lightE}
h(\Lambda;\Lambda) = \frac{\lambda^2}{N^2} \, C_2(\Lambda) \ ,
\end{equation}
where $C_2(\Lambda)=\frac{1}{2}\langle \Lambda, \Lambda + 2\rho \rangle$ is  the quadratic Casimir of $\mathfrak{su}(N)$. Thus, for fixed $\Lambda$ (independent of $N$ and
$k$), the conformal dimension of these states vanishes in the 't~Hooft limit, thus explaining the name 
`light states'. 

However, if we allow
$\Lambda$ to have Dynkin labels that depend on $k$, this conclusion is not necessarily correct. For example, the 
state $(0;{\rm f})$ whose conformal dimension in the 't~Hooft limit equals $h=\tfrac{1}{2}(1-\lambda)$, 
is actually a light state: reintroducing the affine Dynkin label, it corresponds to 
\begin{equation}
(0;{\rm f}) = ([ k;0, \ldots, 0] ; [k; 1, 0, \ldots, 0])\ ,
\end{equation} 
and hence $(0;{\rm f})$ is related by a field redefinition  to the `light state' 
\begin{equation}
\Lambda^+_j = \Lambda^-_j = k\, \delta_j^{N-1} \ . 
\label{}
\end{equation}	
\smallskip
As we shall see in the course of our discussion the conformal dimension of the light states gets up to $h \sim {\cal O}(N)$.

Using the explicit form of the inner product on weight space, the quadratic Casimir equals 
\begin{equation}\label{Casimir}
C_2(\Lambda) = \sum_{i<j} \Lambda_i \Lambda_j \frac{i(N-j)}{N} 
+ \frac{1}{2} \sum_j \Lambda_j^2\, \frac{j(N-j)}{N} 
+ \sum_{j} \Lambda_j \, \frac{j(N-j)}{2} \ ,
\end{equation}
where $\Lambda_j$ are the Dynkin labels. 
For later purposes it will be useful to rewrite this expression in  an alternate way.
The inner product in weight space can be diagonalised by passing to an orthogonal basis. To this end, define
\begin{equation}
r_i = \sum_{j=i}^{N-1} \, \Lambda_i \ , \qquad r_N = 0 \ , 
\label{rowi}
\end{equation}	
which are nothing but the row lengths of the Young diagrams associated with the representation $\Lambda$. 
The total number of boxes in the diagram is given by 
\begin{equation}
B = \sum_{j=1}^{N} \, r_j = \sum_{j=1}^{N-1} \, j\, \Lambda_j \ .
\label{boxes}
\end{equation}	
The reduced row labels are defined by removing a part proportional to the total number of boxes from the row labels,
\begin{equation}
R_i = r_i -\frac{B}{N}  \ ,
\label{rrowi}
\end{equation}	
and it is in terms of these that the inner product is diagonal. Note that $\sum_{i=1}^N\, R_i =0$.
Then we can write 
\begin{equation}
2\, C_2(\Lambda) =\sum_{i=1}^{N} \left(R_i + \frac{N-2\,i +1}{2}\right)^2 - \frac{N(N^2-1)}{12} \ . 
\label{CasimirR}
\end{equation}	
One slightly unsavory part about this expression is that the reduced row labels are not integers. This can, however,
be remedied by noting that one can equivalently write the quadratic Casimir as 
\begin{equation}
2\, C_2(\Lambda) =\sum_{i=1}^{N} \left(r_i + \frac{N-2\,i +1}{2}\right)^2 -\frac{B^2}{N}- \frac{N(N^2-1)}{12} \ ,
\label{CasimirSq}
\end{equation}	
where the entries in the bracket 
\begin{equation}
n_i \equiv r_i + \frac{N-2\,i +1}{2} 
\label{nidef}
\end{equation}	
are (half-)integral. These (half-)integers are constrained to lie in a band set by $N$ and $k$, viz.,
\begin{equation}
k+\frac{N-1}{2} \geq n_1 > n_2 > \cdots >n_N = \frac{1-N}{2} \ .
\label{niconstraint}
\end{equation}	

\subsection{The Number of Light States}

To begin our discussion, let us obtain an estimate for the total 
number of light primaries as a function of $N$ and $k$. As we have
seen above, the light states are uniquely characterised in terms of a representation $\Lambda\equiv \Lambda^+$
of $\mathfrak{su}(N)_k$. The corresponding Dynkin labels $\Lambda_i$ define an 
$(N-1)$-tuple with the constraint \eqref{lampmk}. To count the total number of light states, we consider first the 
representations at a fixed level $\sum_j \Lambda_j = l \leq k$.  Since different $\Lambda_i$ correspond to different representations,
the light states at level $l$ are counted by the 
{\em weak compositions} of $l$ into $N-1$ parts.\footnote{We recall that the composition of an integer $M$ is the 
ordered set of integers which sum to $M$, i.e., we consider permutations of integer partitions. A weak composition further 
allows the sum to contain zeros.}

For fixed $l$ the number of weak compositions is precisely given as the binomial coefficient 
${l + N-2 \choose l}$, as can be inferred by realising that we have to distribute $(N-1) -1$ screens between 
$l$ items. Summing over all possible values of $l \leq k$ we end up with the number of light states 
${\cal N}_{N,\lambda}$ 
\begin{equation}
{\cal N}_{N,\lambda} = {N+k-1  \choose k} \ .
\label{Nlambda}
\end{equation}	
In the 't Hooft scaling limit we therefore have an exponentially large number of light states
\begin{equation}
{\cal N}_{N,\lambda} \sim e^{N\, G(\lambda)} \ , \qquad 
G(\lambda) = -\left[\log (\lambda) + \frac{1-\lambda}{\lambda}\, \log\left(1-\lambda\right) \right] \ .
\label{Glam}
\end{equation}

We are interested in understanding the distribution of the conformal weights $h(\Lambda;\Lambda)$, as we scan over all the light states. The rationale for focussing on this particular subset of primaries is that for low enough conformal dimensions these are the only states in the system; one might thus expect that their distribution controls the thermodynamic behaviour of the partition function at very low temperatures.

\section{Numerical Results}
\label{sec:numerics}

The distribution of light states which is of prime interest becomes somewhat easier to intuit once one can visualise the general nature of the spectrum. In order to get a feeling for the distribution of the light states, we have done some numerical simulations, enumerating the number of states with a fixed conformal dimension. We first describe the results of our numerical experiments in this section. In the subsequent sections, we shall then also explain how some of these results can be obtained analytically.

In principle the determination of the spectrum of light states is straightforward: one lists all the representations and computes the conformal dimensions from the quadratic Casimir using \eqref{lightE}.  The only problem we encounter is that
the number of light states grows exponentially, cf., (\ref{Glam}). Thus, we can only make a full analysis for small values of $N$ and $k$.\footnote{We have found that we can reasonably compute data for combinations of $N$ and $k$ such that the total number of states ${\cal N}_{N,k} \lesssim 10^8$.}

However, for larger values of $N$ and $k$ one can still sample the spectrum of light states
quite successfully. To this end one first picks a level $l\in \mathbb{N}$, using the distribution
\be\label{leveldist}
p_{N,k}(l) = \frac{{N+l -2\choose l}}{{N+k-1\choose k}} = \text{BetaBinomial}(N-1,1,k;l) \ ,
\ee
and then chooses a random Young diagram at this level. For each of these Young diagrams one then calculates 
the corresponding conformal dimension using (\ref{lightE}). This random sampling technique allows us to access the typical states in the distribution and gives an accurate portrait for such states. We find that we can get extremely reliable results by sampling about a million random tableaux. Clearly, the statistics of the distribution works in our favour  and one thus obtains a good estimate for the conformal weight  distribution of the light states (away from the tail).

In the sequel we will describe the results of numerical experiments both from the complete spectrum, and from the random sampling technique, and the lessons one can draw regarding the spectrum of light states.

\subsection{The Maximal Conformal Dimension}\label{s:3.1}

\begin{figure}[h]
\begin{center}
\includegraphics[width=3.5in]{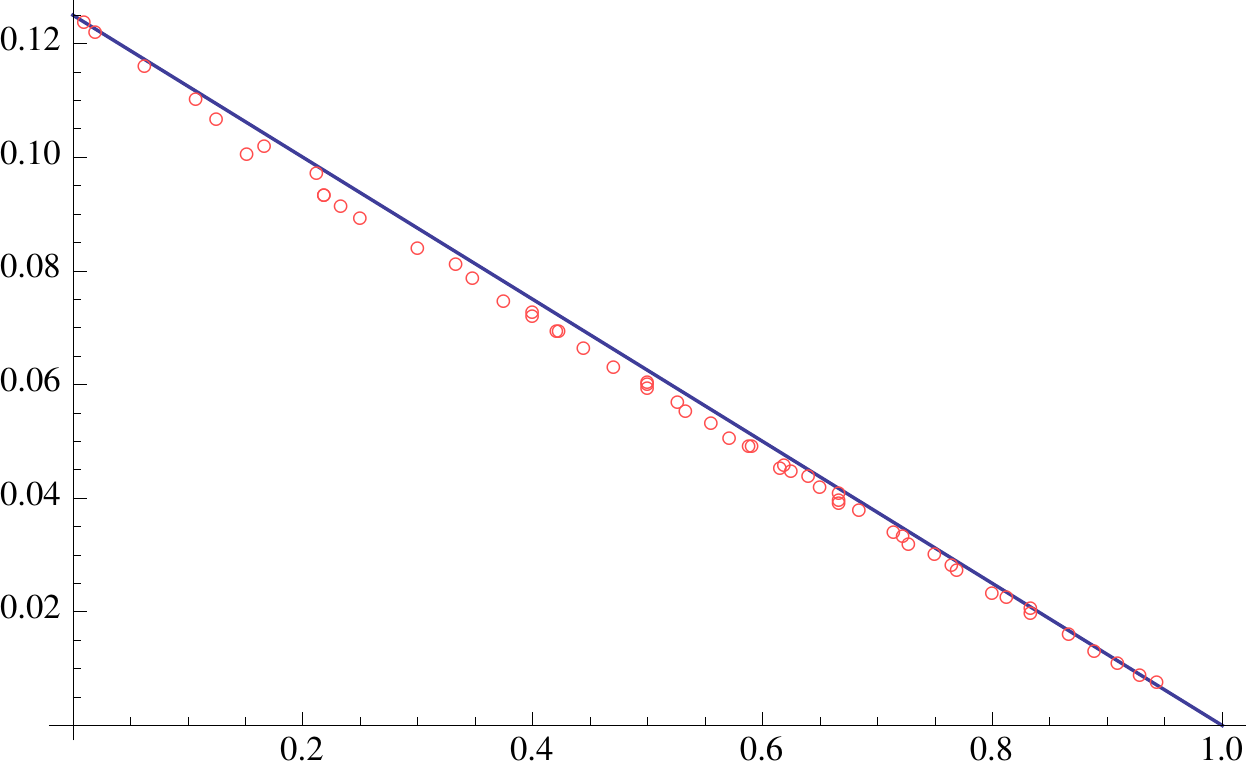}
\begin{picture}(0,0)
\setlength{\unitlength}{1cm}
\put (0,0) {$\lambda$}
\put(-10,5){$\frac{\hmax}{N}$}
\end{picture}
  \caption{Numerical result (red circles) versus analytic fit \eqref{lhmax} (blue line)   for the maximum conformal dimension (normalized by $N$ to facilitate
 comparison) attained for given $N$ and $\lambda$.  We obtained the full spectrum of light states for discrete values of  $N \in \{2,3,\cdots,50 \} $ and $k \in \{1,2,\cdots, 200 \}$, restricting to situations where the total number of light states is less than $10^7$. We also note that the representation attaining $\hmax$ agrees excellently with \eqref{lmaxOm}.}
\label{f:lhmax}
\end{center}
\end{figure}
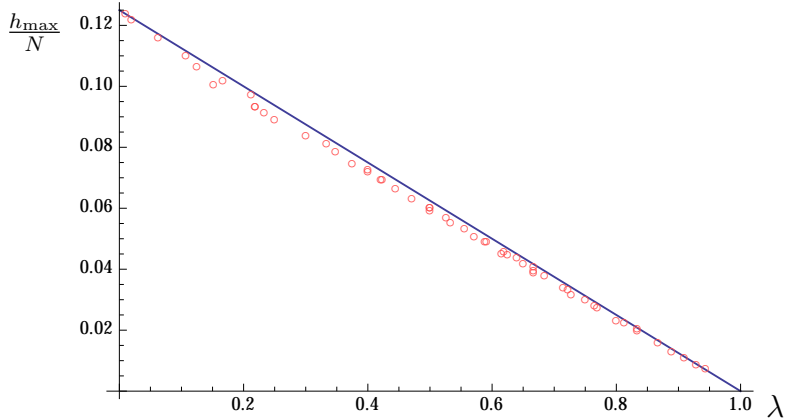

Since there are only finitely many light states (for given $N$ and $k$) the conformal dimensions are bounded from below
and above. The lower bound is obviously $0\leq h$, and our numerical results suggest that the upper bound is 
\be\label{lhmax}
\hmax(N,k) = \frac{N}{8} (1-\lambda) \ . 
\ee
The behaviour of the maximal conformal dimension for a choice of values of $2\leq N \leq 50$ and $1\leq k \leq 200$  is illustrated in \fig{f:lhmax}. 

Incidentally, this maximum value is approximately attained by the representation 
\be\label{lmaxOm}
\Lambda^{\rm m}_j =  k \; \delta_{j, [N/2]}  \ ,
\ee
where $[x]$ denotes the integral part of $x$. Its conformal dimension can be easily estimated to be 
\begin{equation}
h(\Lambda^{\rm m};\Lambda^{\rm m}) = \frac{1}{p (p+1)} C_2(\Lambda^{\rm m}) = \frac{1}{2\,p\, (p+1)} 
\,k\, \frac{N^2}{4} \, \left(\frac{k}{N} +1 \right)  \approx \frac{N}{8} \, (1-\lambda)\ ,
\label{lbest}
\end{equation}
where we have used \eqref{Casimir} as well as $\tfrac{N}{p}=\lambda$ and $\tfrac{k}{p}=1-\lambda$. 
Note that since there are very few states near the maximum, it is not possible to obtain a good estimate for it based on our sampling technique. We have therefore only tested this prediction for relatively small values of $N$ and $k$ specified above. 

\subsection{The Peak of the Distribution}\label{sec:peak}

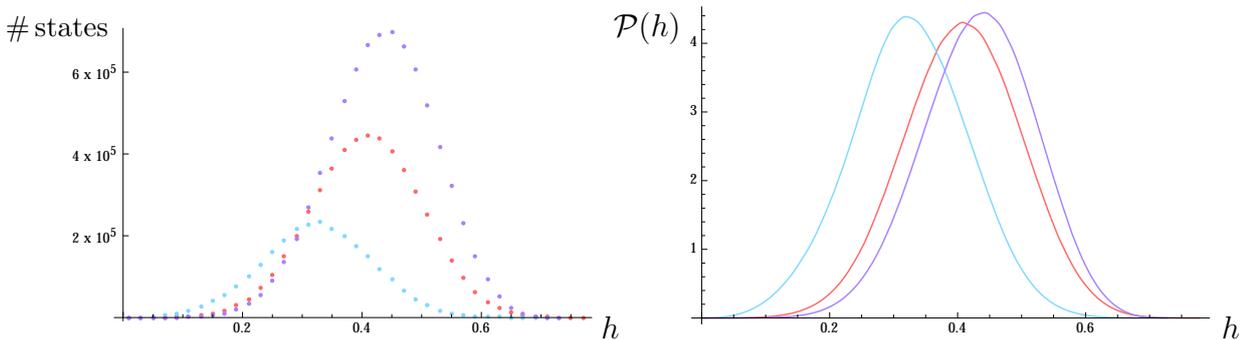
\begin{figure}[h]
\begin{center}
\includegraphics[width=2.7in]{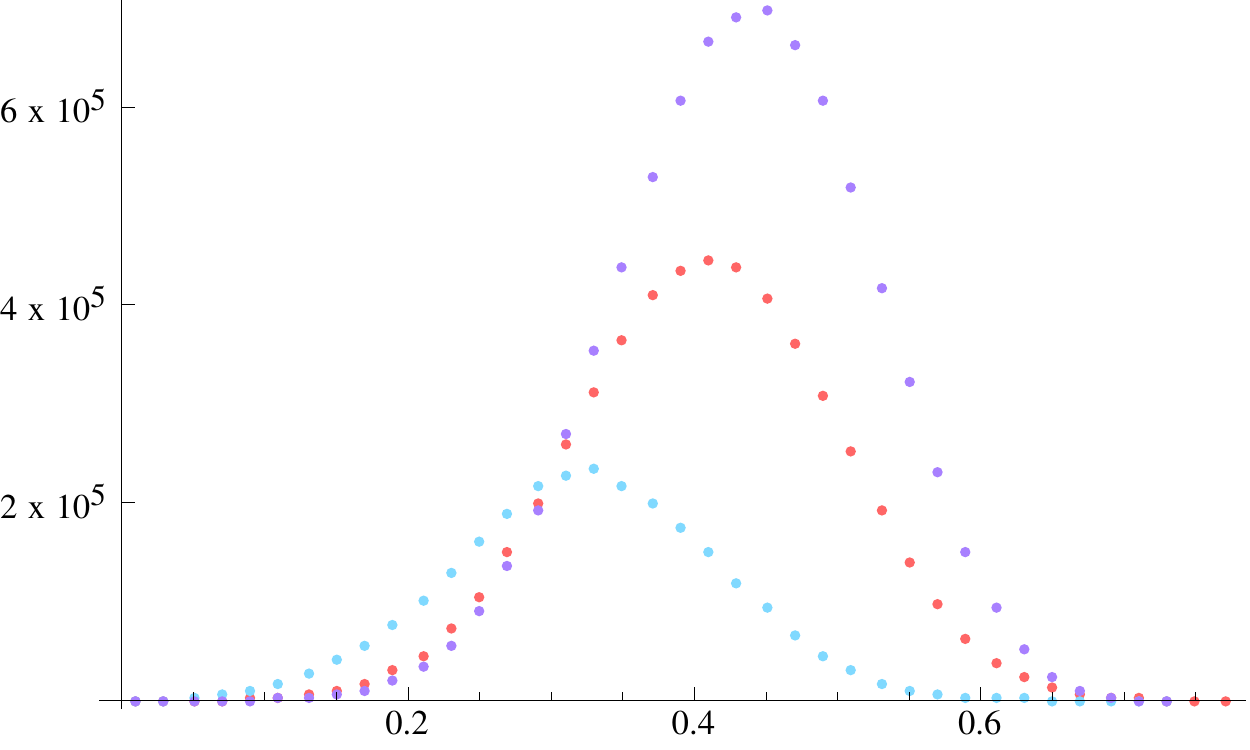} 
\hspace{1cm}
\includegraphics[width=2.7in]{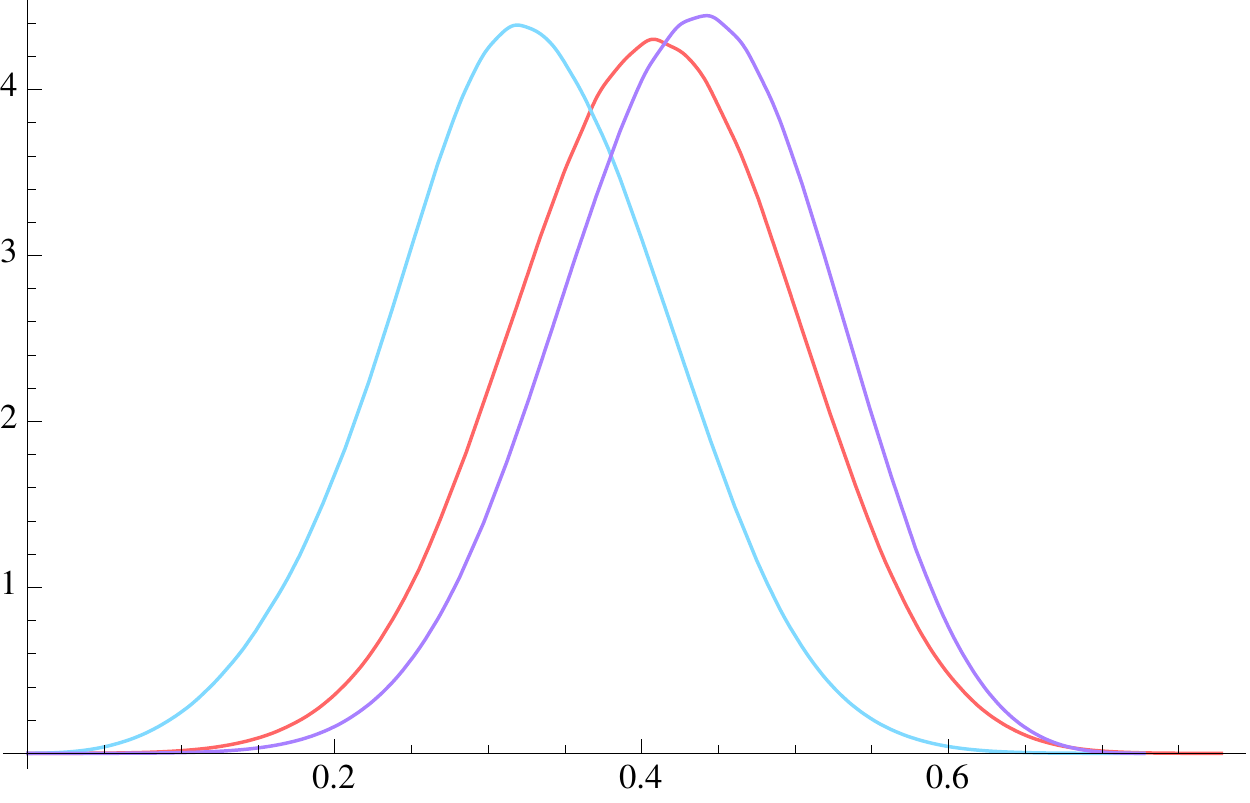} 
\begin{picture}(0,0)
\setlength{\unitlength}{1cm}
\put (-16,4) {{\small \#\,\text{states}}}
\put (-8.3,0) { $h$}
\put (-8,4) {${\cal P}(h)$}
\put (0,0) {$h$}
\end{picture}
\caption{Distribution of the conformal dimensions for light states. The left panel shows the histogram of the distribution and the right panel shows a fit to a smooth probability distribution (see text for details). Color coding:  $N=8,k=24$ (blue), $N=k=13$ (red), and $N=24,k=8$ (purple).
} 
\label{f:lightdistributions}
\end{center}
\end{figure}

Having delineated the bounds on the spectrum, we now turn towards understanding the degeneracies of states. 
Clearly, there is a unique vacuum state with $h=0$, and as we mentioned above few states near the maximum.
At the same time it is clear from the expression \eqref{lightE} that different 
representations can have approximately the same conformal dimension. To get a feeling for the distribution of the
spectrum we display plots of the number of states against the conformal dimension for some sample values of 
$N$ and $k$ in \fig{f:lightdistributions}.

The histogram of the conformal dimensions in \fig{f:lightdistributions} is shown as a bin-count over $50$ bins spanning the interval $[0,\hmax]$. To convert this data into a probability distribution, we use a kernel smoothing technique. This involves picking a bandwidth $\delta$ (bin size) and a smoothing kernel $K(x)$ which is a symmetric function which integrates to $1$; we have chosen $K$ to be a Gaussian. The normalised probability distribution function is then given as 
\begin{equation}
{\cal P}(x) = \frac{1}{n\, \delta} \, \sum_{i=1}^n\, K\left(\frac{x-x_i}{\delta}\right) \ ,
\label{skdist}
\end{equation}	
where $x_i$, $i = 1, \ldots, n$ are the data-points, i.e., the bin counts in the histogram plot of our sample. 

There are two key features of the spectrum that we wish to highlight.
Firstly, the spectrum has a reflection symmetry under $\lambda \to 1-\lambda$, where 
we flip simultaneously the probability distribution under $\bar{h} \to 1-\bar{h}$ (with $\bar{h}=\frac{h}{h_{\rm max}}$).   
This will become more evident
below, see \fig{f:lam01plots}, and it has its origins in the level-rank duality $k \leftrightarrow N$
of the coset minimal models (see the discussion around \eqref{C2rel}).  Secondly, we see from \fig{f:lightdistributions}
that  the degeneracy of states has a characteristic peak at a particular value of the conformal 
dimension.\footnote{Small values of $N$ in the limit $\lambda \to 0$
and small values of $k$ in the limit $\lambda \to 1$ are an exception to this statement, as might be expected since we are away from the 't Hooft scaling regime.  As we will also see later, the 
$\lambda \to 0,1$ limits, the former of  which has been discussed earlier in \cite{Gaberdiel:2011aa,Banerjee:2012aj},  are somewhat 
degenerate limits.}
 Based on our numerical results (as well as the analytical treatment in \S\ref{sec:ff}) it seems that the peak of the distribution occurs for the conformal dimension
\be\label{peak}
\hpeak = \frac{N}{24} (1-\lambda^2) = \frac{c}{24} \ .
\ee
Let us first understand the peak in the distribution and then return to the shape of the distribution.

The problem of finding the peak of the distribution is equivalent to ascertaining the set of diagrams that are typical 
with respect to the measure given by the quadratic Casimir. While typical representations have been encountered before
for various other measures, see e.g., \cite{Vershik:1996uq,Vershik:2009kx},  as far as we are aware the problem at hand has not been 
addressed in the literature. 

\begin{figure}[h!]
\begin{center}
\includegraphics[width=6in]{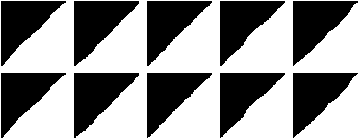}
\caption{
Pictorial view of a random set of 10  Young diagrams representing states near the peak of the randomly generated distributions for $N=k=500$. The diagrams are mostly triangular in this asymptotic limit. Note that we have sampled $10^6$ random Young diagrams, which 
is a very small fraction of the total number of states (which is $\approx 10^{299}$)  but apparently suffices to see the typical diagram. } 
\label{f:ytlightNk500}
\end{center}
\end{figure}

It is however easy to intuitively motivate (\ref{peak}): in the 't~Hooft limit, the set of representations which lead to the 
typical conformal dimension $\hpeak$ should be such that addition or removal of ${\cal O}(1)$ boxes does not 
modify the quadratic Casimir by a large amount. So to understand where most of the states in the spectrum lie, we simply 
need to ask what shapes of diagrams allow maximal number of ${\cal O}(1)$ deformations. Clearly, the best case is when 
we have half of the $k\times N$ rectangle filled with boxes (which can then be removed or augmented). For thinner or thicker 
diagrams, on the other hand, the constraints from the edges come into play.
Let us therefore consider a roughly triangular Young diagram with Dynkin labels
\begin{equation}
\Lambda^{\rm p}_i = \frac{k}{N} \ .
\end{equation}
Using (\ref{Casimir}) it is not hard to calculate the corresponding value for the quadratic Casimir 
as
\begin{equation}
C_2 (\Lambda^{\rm p})  = \frac{(N^2-1)}{24} \Bigl[ \frac{k^2}{N} + 2 k \Bigr]  \quad
\Longrightarrow \quad
h(\Lambda^{\rm p}; \Lambda^{\rm p}) = \frac{C_2(\Lambda^{\rm p})}{p (p+1)} \approx \frac{N}{24} (1-\lambda^2) \ .
\end{equation}
This interpretation is also supported by our numerical results. For $N=k=500$ we exhibit
in \fig{f:ytlightNk500} a representative set of Young diagrams near the peak; the results confirm very nicely the above picture. 

We will also see in \S\ref{sec:ff} an analytical demonstration of these facts using the free fermion picture.

%

\subsection{The Shape of the Distribution}
\label{sec:shape}
We have already noted that the spectrum demonstrates a reflection symmetry about $\lambda = \half$ due to the level-rank duality. 
Indeed, from $\hpeak/\hmax = \frac{1+\lambda}{3}$ we see that the spectrum goes from being positively skewed at $\lambda =0$ to negatively skewed at $\lambda =1$.  Despite this lop-sidedness in the spectrum for $\lambda\neq \half$, it is easy to see that  the distribution of the light states is well approximated by the normal  distribution 
\begin{equation}
{\cal D}(\barh)   = {\cal A}_{\barh} \, \exp\left(-\frac{ ( \barh - \barh_{\rm mp})^2}{2\,\sigma_{\barh}^2}\right) ,
\label{Dlight}
\end{equation}	
for a wide range of $N$ and $\lambda$.  Here we have introduced the rescaled conformal dimension $\barh$, and the mean and variance are approximately 
given as 
\be
\barh = \frac{h}{\hmax} \ , \qquad
\barh_{\rm mp} = \frac{\hpeak}{\hmax} = \frac{1}{3} (1+\lambda) \ , \qquad
\sigma_{\barh} \propto \frac{1}{\sqrt{N}} \ .
\ee
Furthermore, the normalisation constant $ {\cal A}_{\barh}$ is determined by the requirement that 
\be
\int_0^1\, d\barh \, {\cal D}({\barh}) = {\cal N}_{N,\lambda}\ ,
\ee
where ${\cal N}_{N,\lambda}$ is given in  (\ref{Nlambda}).

\begin{figure}[t!]
\begin{center}
\includegraphics[width=1.82in]{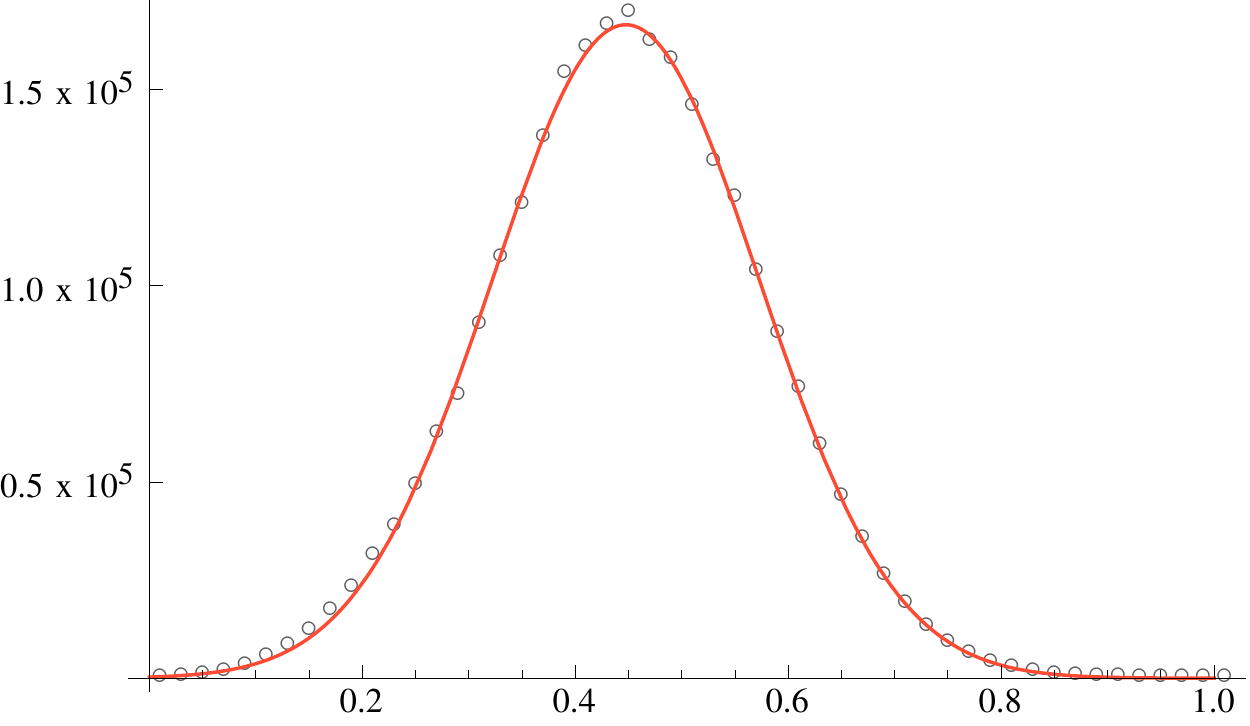}
\hspace{3mm}
\includegraphics[width=1.82in]{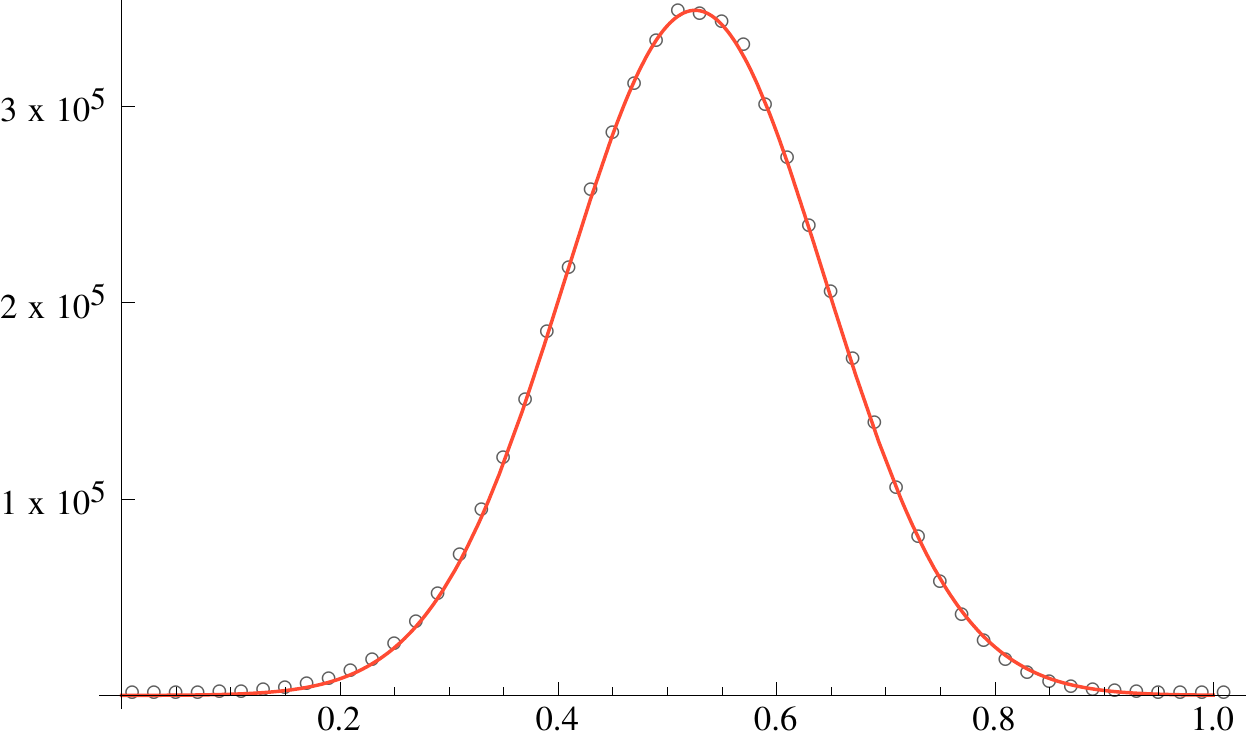}
\hspace{3mm}
\includegraphics[width=1.82in]{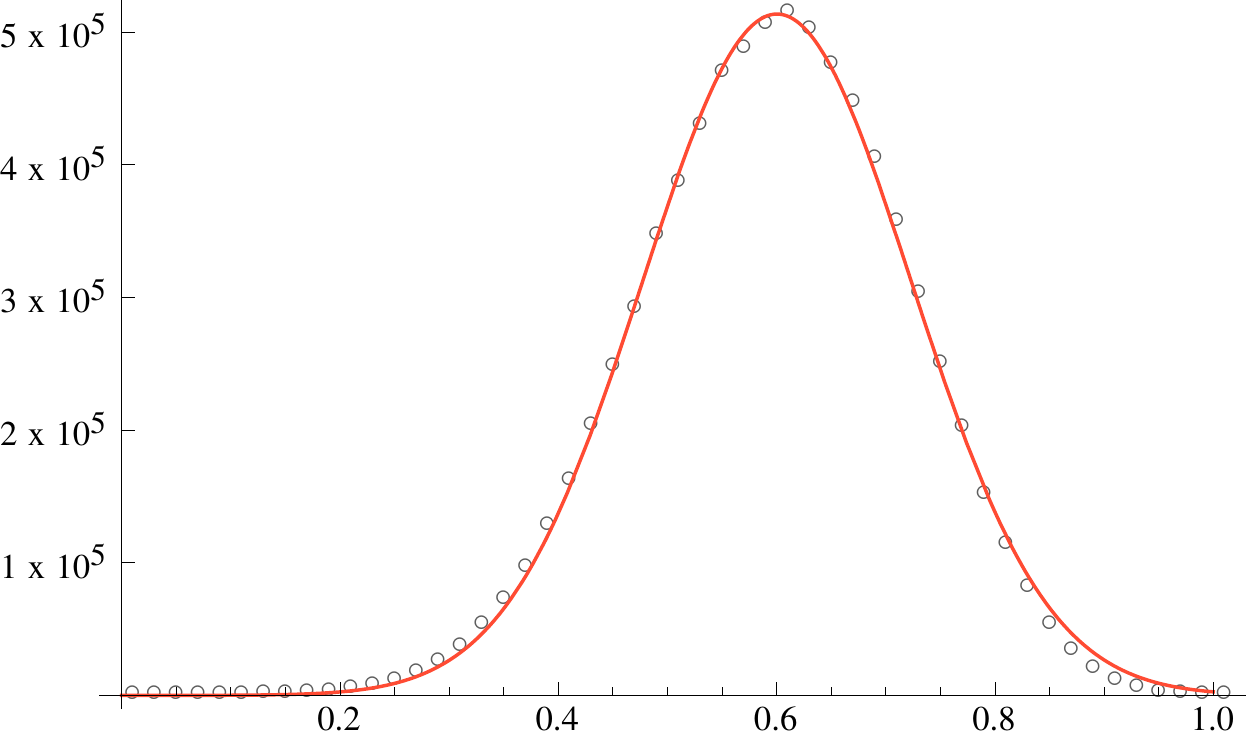}
\begin{picture}(0,0)
\setlength{\unitlength}{1cm}
\put (-13.6,-0.3) { {\scriptsize $N=8, k = 24$}}
\put (-8.3,-0.3) { {\scriptsize $N= k = 13$}}
\put (-3.0,-0.3) { {\scriptsize $N=24, k = 8$}}
\put (-15.7,2.5) {{\scriptsize ${\cal D}(\barh)$}}
\put (-10.4,0) {{\scriptsize $\barh$}}
\put (-10.6,2.5) {{\scriptsize ${\cal D}(\barh)$}}
\put (-5.2,0) {{\scriptsize $\barh$}}
\put (-5.4,2.5) {{\scriptsize ${\cal D}(\barh)$}}
\put (0,0) {{\scriptsize $\barh$}}
\end{picture}
\caption{The distribution of primary scaling dimensions for the full spectrum obtained for different values of $N$ and $k$. The dots represent the bin counts of the histogram data and the solid curve is a fit to the Gaussian distribution. We present the same choices of $N$ and $k$ as in \fig{f:lightdistributions}.} 
\label{f:lighthnormal}
\end{center}
\end{figure}
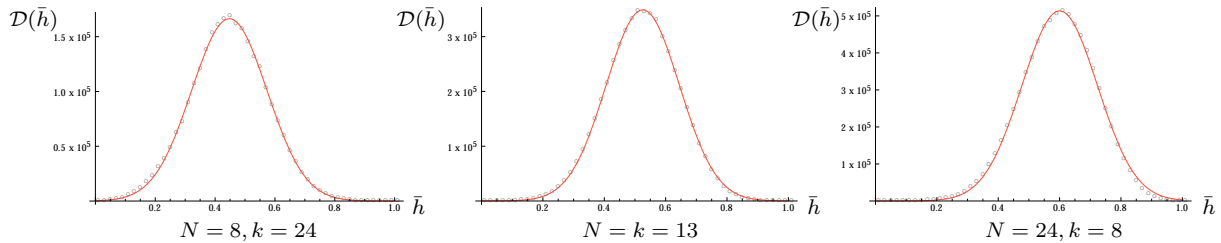

The main evidence for this claim comes from our numerical explorations. In particular, we have, for 
a variety of different values of $N$ and $k$, sampled the light spectrum. We have binned the rescaled conformal dimension 
$\barh$ into a set of bins spanning the unit interval, and have taken a count of the number of states in a given bin. The
 resulting histogram data was then fitted to a normal distribution. For some sample values of $N$ and $k$ the result is 
 displayed in \fig{f:lighthnormal}. As one can see even for relatively small values of $N, k \sim {\cal O}(10)$ 
 (and with $\lambda$ away from $0,1$) one has a
 pretty good fit to the normal distribution.

The data obtained using the random sampling algorithm allows us to find numerical fits to the mean and the variance of the Gaussian distribution. Numerically we find the best fit for the parameters  
\begin{align}
\barh_\text{mp}\,\big|_\text{fit} & = 0.317 \, \lambda + 0.343  \nonumber \\
\sigma^2_{\barh}\,\big|_\text{fit} & = \frac{1}{10^2\, N^{0.979}} \left(0.664 + \frac{7.64}{1- \lambda} + \frac{0.014}{\lambda} \right)
\label{fitsNlammeanvar}
\end{align}
which supports our assertion above. These fits are based on data sets involving 
$N \in [10,100]$ and $k \in [10,1000]$.

We can also consider the corresponding canonical ensemble, defined by 
\begin{equation}\label{numcan}
Z(\beta) = \int_{0}^1\, d\barh\, e^{-\beta \,\hmax\, \barh }\;{\cal D}(\barh)  \ .
\end{equation}
Since both $\hmax$ and $\frac{1}{\sigma_\barh^2}$ (in the gaussian fit \eqref{Dlight}) are ${\cal O}(N)$ we only 
need to take the saddle value of $\barh$ in \eqref{numcan}. This saddle value occurs for 
$\barh_{\rm sad}=\barh_\text{mp} - c(\lambda)\,\beta$, where $c(\lambda)$ is determined by the 
$\lambda$ dependence of $\sigma_\barh$ and $\barh_\text{mp}$.  We can trust this expression as long as 
$\barh_{\rm sad}$ is sufficiently far away from the tails of the distribution. 

The fact that $\barh_{\rm sad}$ is a continuous function
of the temperature indicates that the dominant representation shifts slowly from near zero at low temperatures to $\barh_\text{mp}$ at high temperatures. This is unlike the jump in saddle point that one sees in the case of, say, the D1-D5 CFT. This is an indication of a lack of phase transition, at least in the light state sector.

\section{CFT Analysis for $\lambda\approx 0$ and $\lambda\approx 1$}
\label{sec:lam01}

As we have just seen, the spectrum of the light states discussed above is mostly featureless (being well approximated by a normal distribution). There are however two degenerate limits $\lambda \to 0,1$, as evidenced from \eqref{Glam}, which deserve separate treatment. In fact, for $\lambda\approx 0$ and $\lambda\approx 1$ we can use a combination of heuristics and a description of the CFT in terms of free fermions/bosons 
to estimate the distribution of the light states in particular for very small conformal dimensions.  Let us first consider the case $\lambda\approx 0$ that
was already studied in \cite{Banerjee:2012aj}, and then turn to $\lambda \approx 1$.

\subsection{The Situation near $\lambda\approx 0$}
\label{sec:lam0}

\begin{figure}[h]
\begin{center}
\includegraphics[width=2.4in]{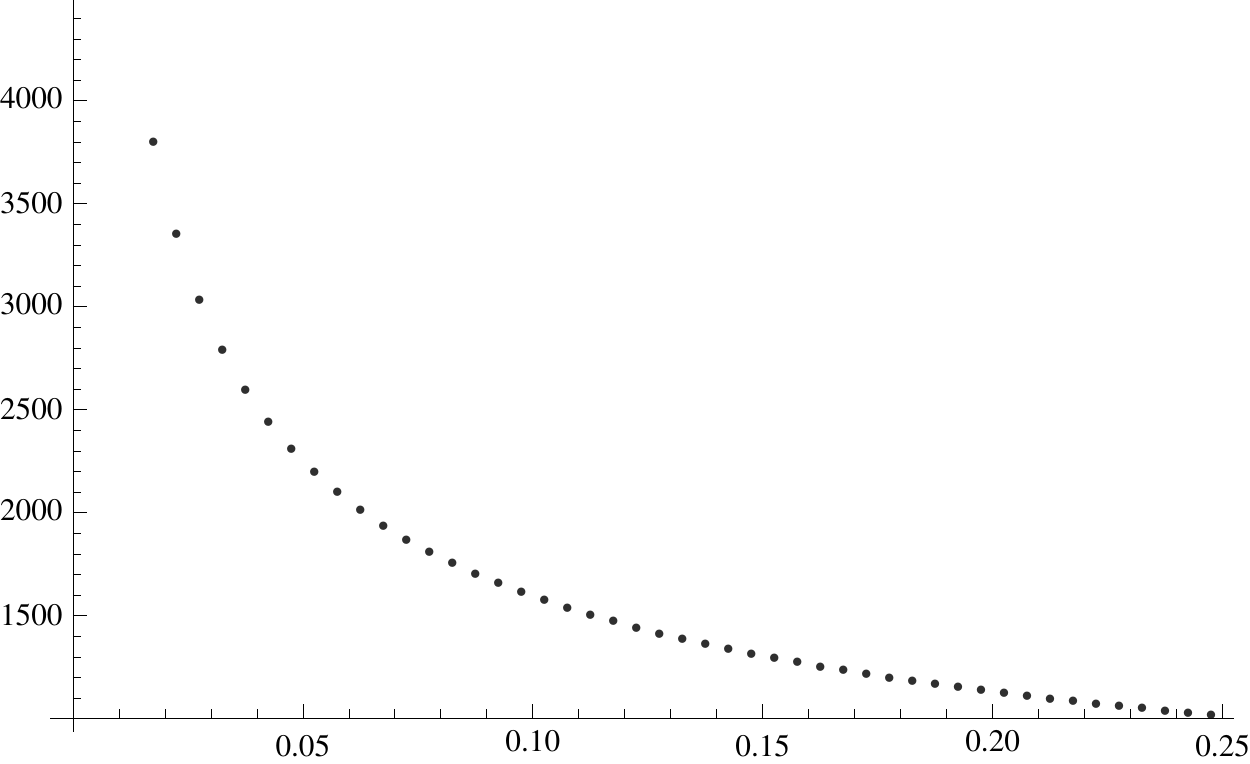}
\hspace{10mm}
\includegraphics[width=2.4in]{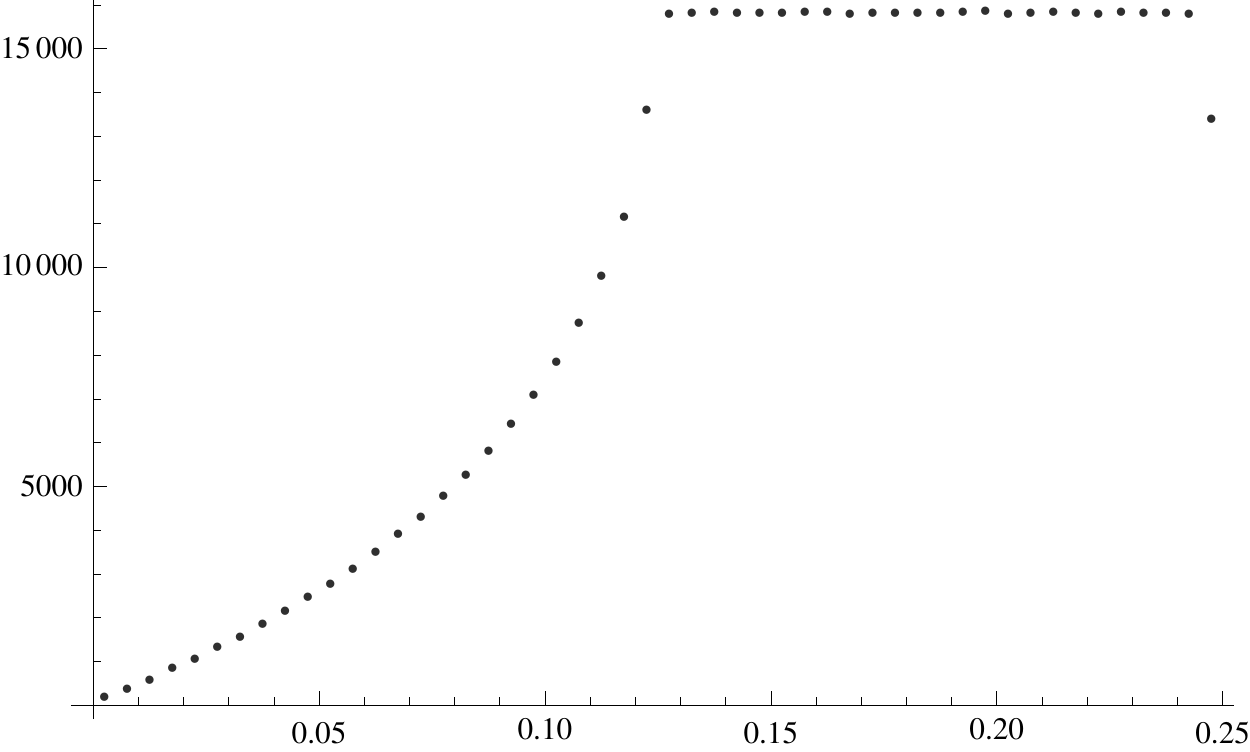}
\\  
\bigskip
\includegraphics[width=2.4in]{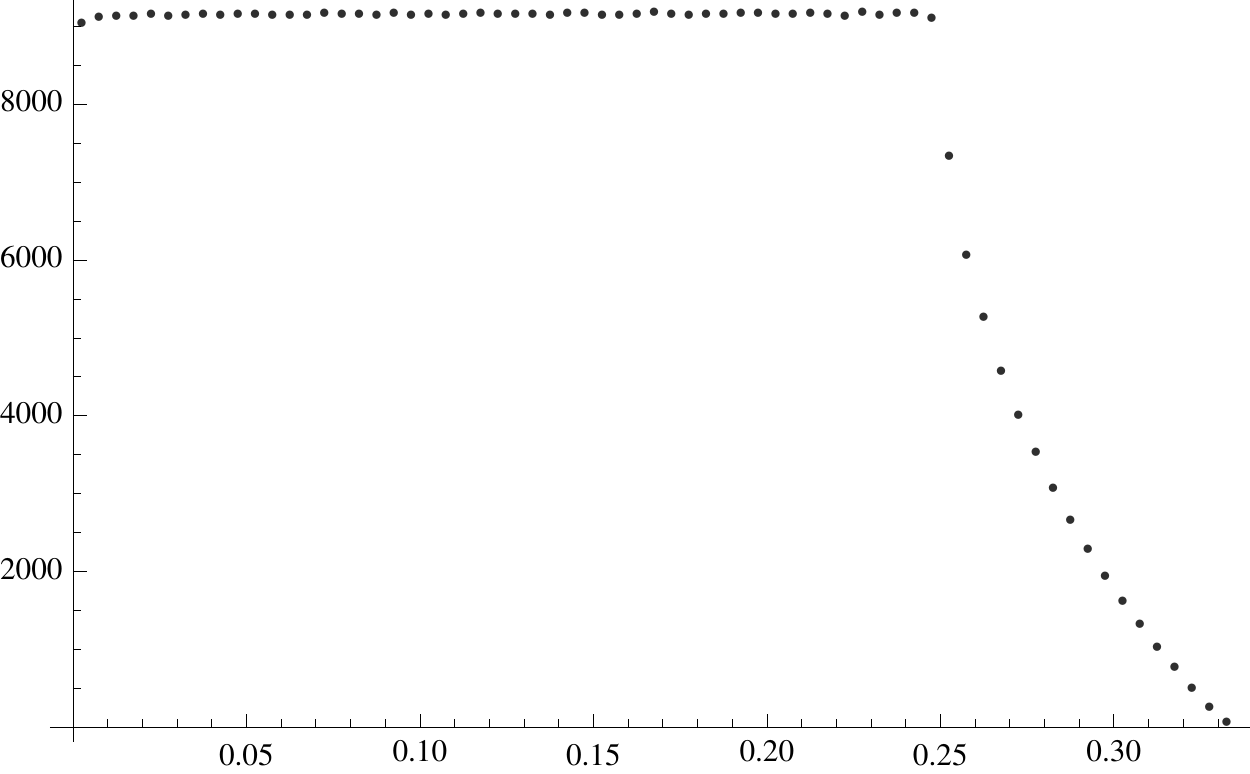}
\hspace{10mm}
\includegraphics[width=2.4in]{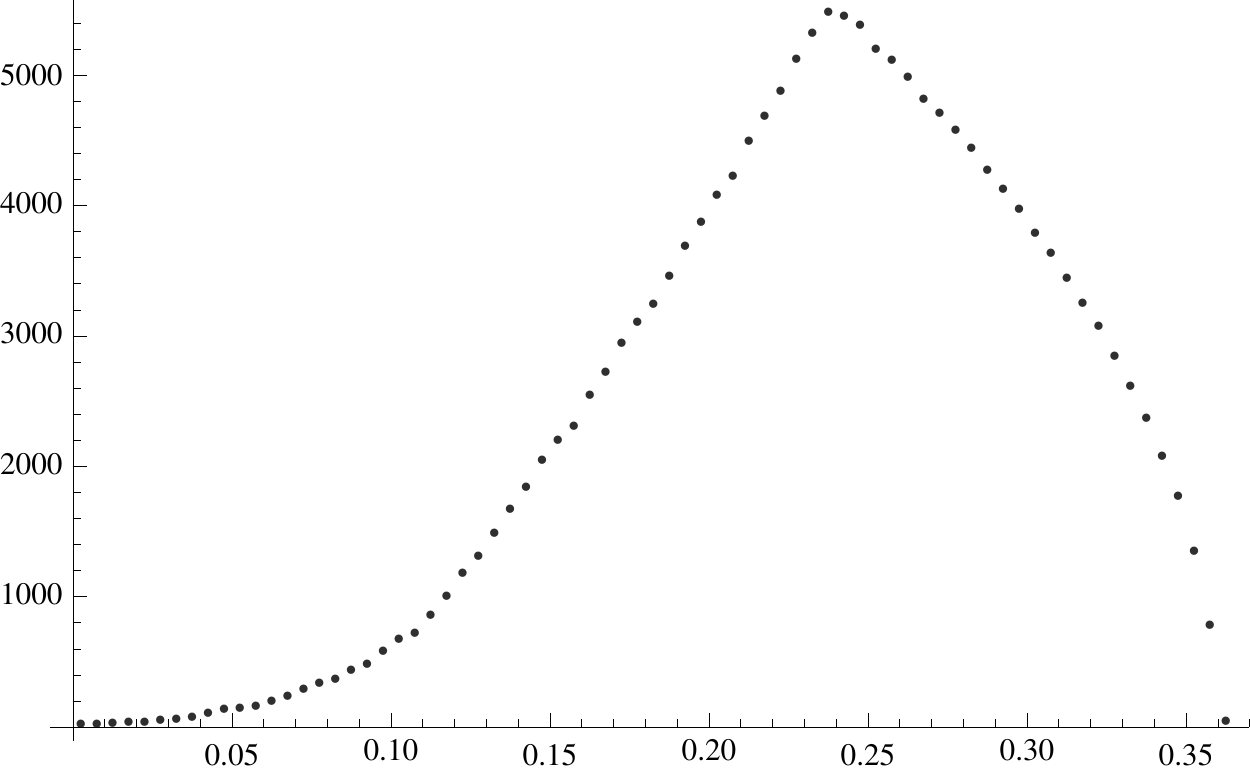}
\\  
\bigskip
\includegraphics[width=2.4in]{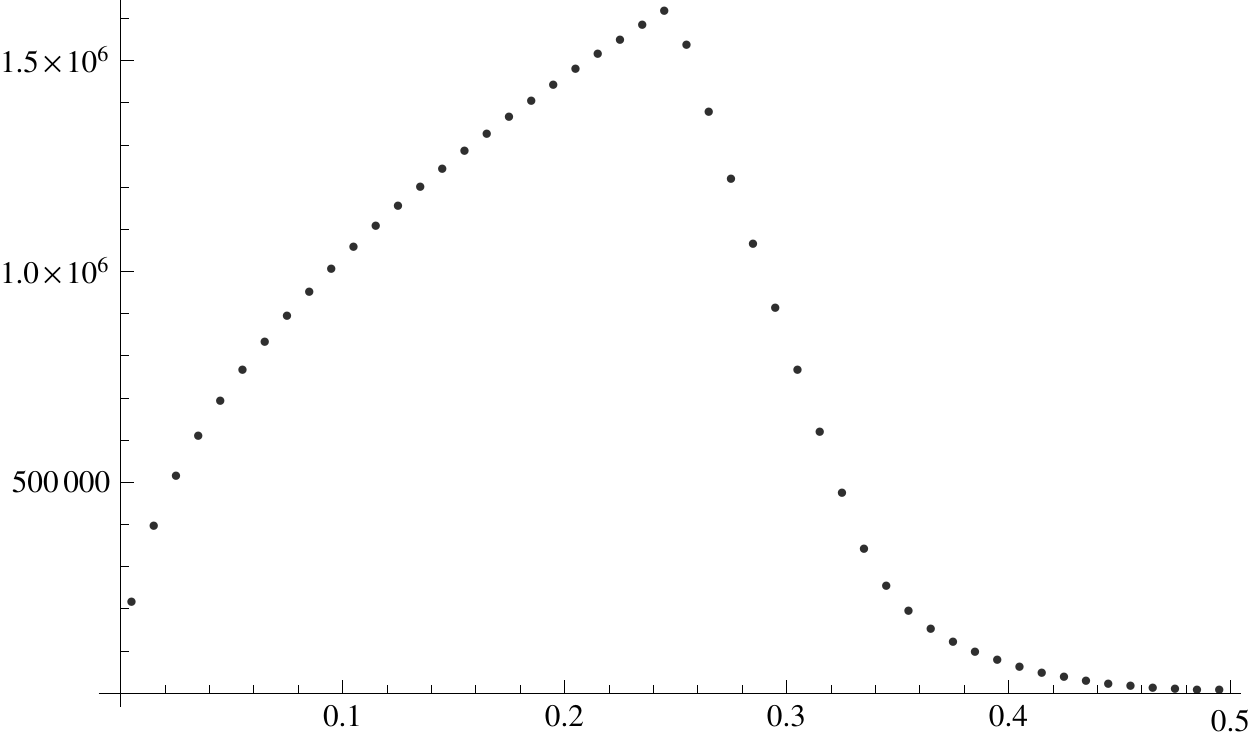}
\hspace{10mm}
\includegraphics[width=2.4in]{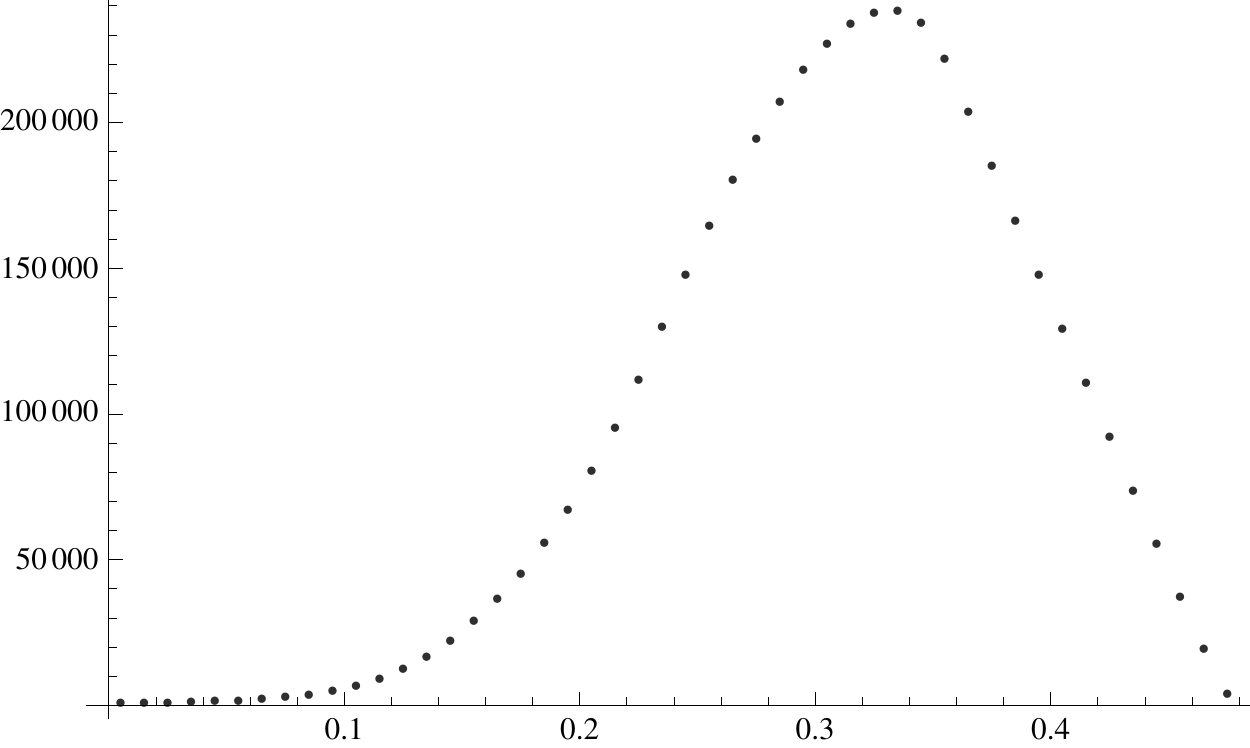}
\\  
\bigskip
\includegraphics[width=2.4in]{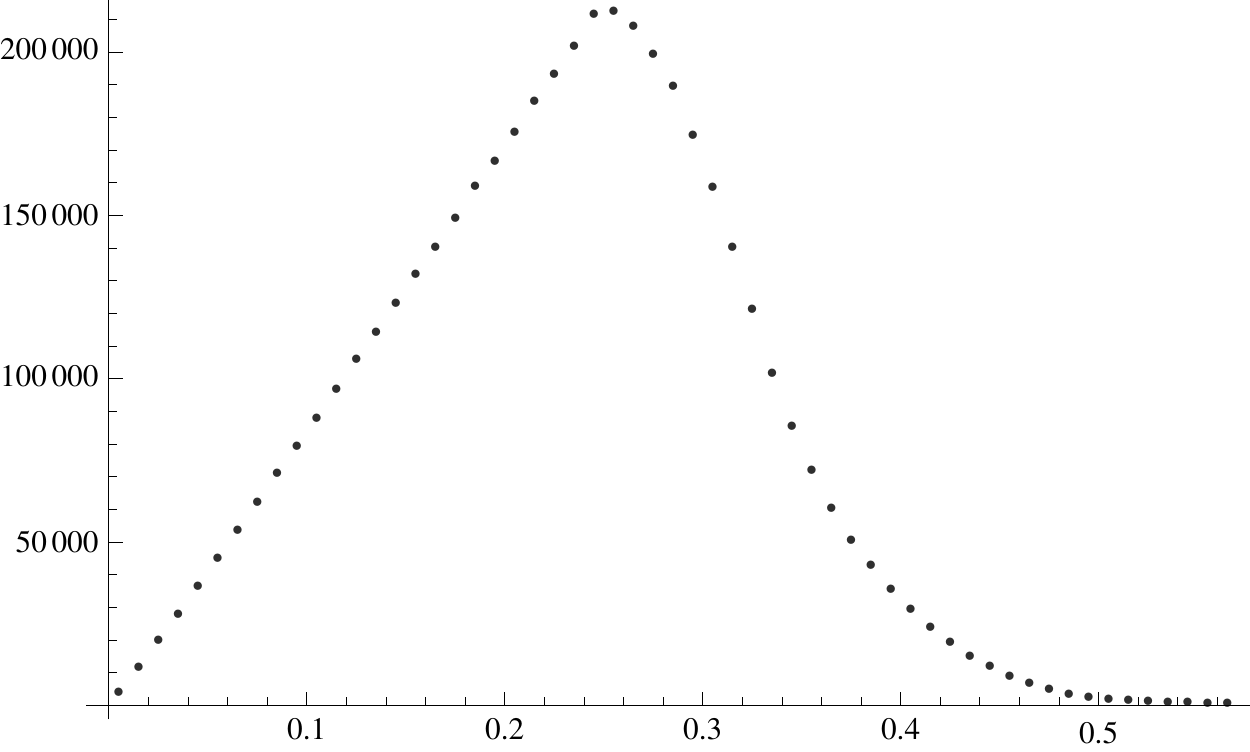}
\hspace{10mm}
\includegraphics[width=2.4in]{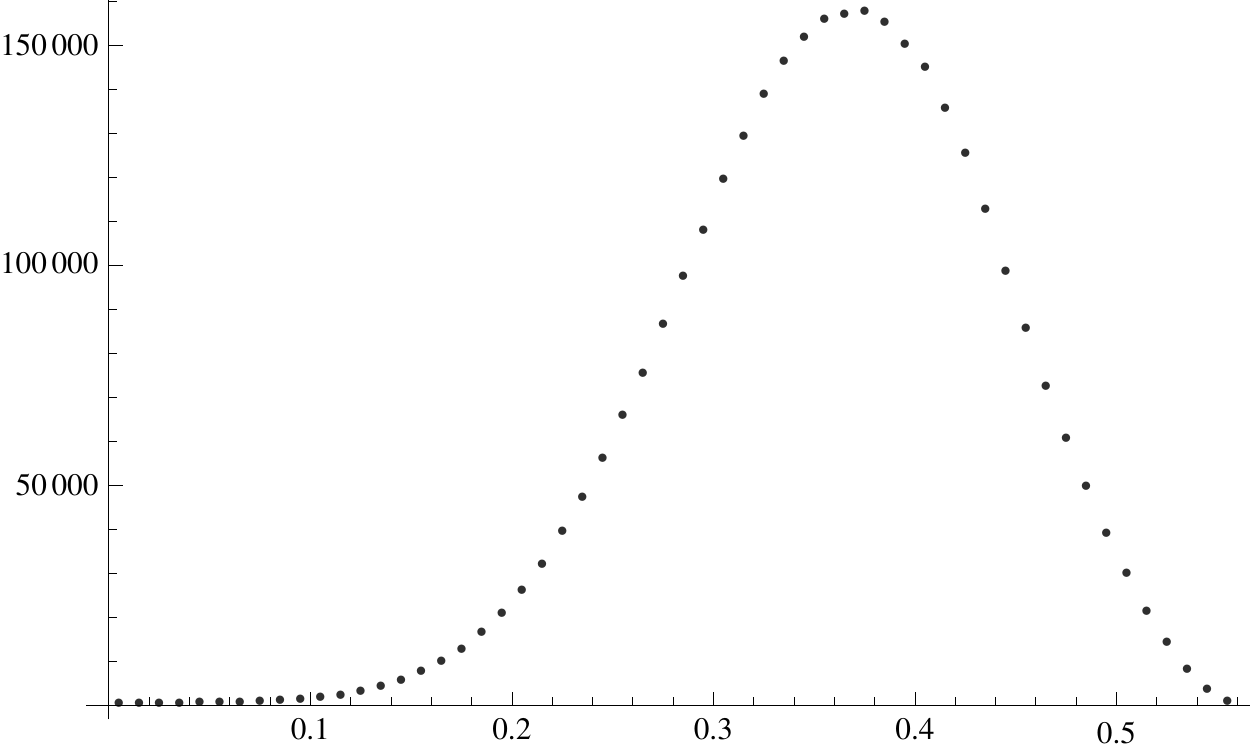}
\begin{picture}(0,0)
\setlength{\unitlength}{1cm}
\put(-11,14.5) {{\scriptsize $N=2$, $k=10^5$}}
\put(-3,14.5) {{\scriptsize $N=10^3$, $k=2$}}
\put(-11,10) {{\scriptsize $N=3$, $k=10^3$}}
\put(-3,10) {{\scriptsize $N=100$, $k=3$}}
\put(-11,5.5) {{\scriptsize $N=4$, $k=600$}}
\put(-3,5.5) {{\scriptsize $N=100$, $k=4$}}
\put(-11,1.5) {{\scriptsize $N=5$, $k=100$}}
\put(-3,1.5) {{\scriptsize $N=50$, $k=5$}}
\end{picture}
\caption{Distribution of light states in the regimes $\lambda \approx 0$ (left) and $\lambda \approx 1$ (right). We draw attention to two facts: (i) the number of states with dimension $h \ll 1$ grows as a power law ${\cal D}(h) \sim h^\alpha$ with $\alpha_{\lambda \approx 0} = \frac{N-3}{2}$ and $\alpha_{\lambda\approx 1} = k-1$ and (ii) the level-rank duality is clearly visible with $N - 1 \leftrightarrow k$ (see \S\ref{sec:lam1}).}
\label{f:lam01plots}
\end{center}
\end{figure}
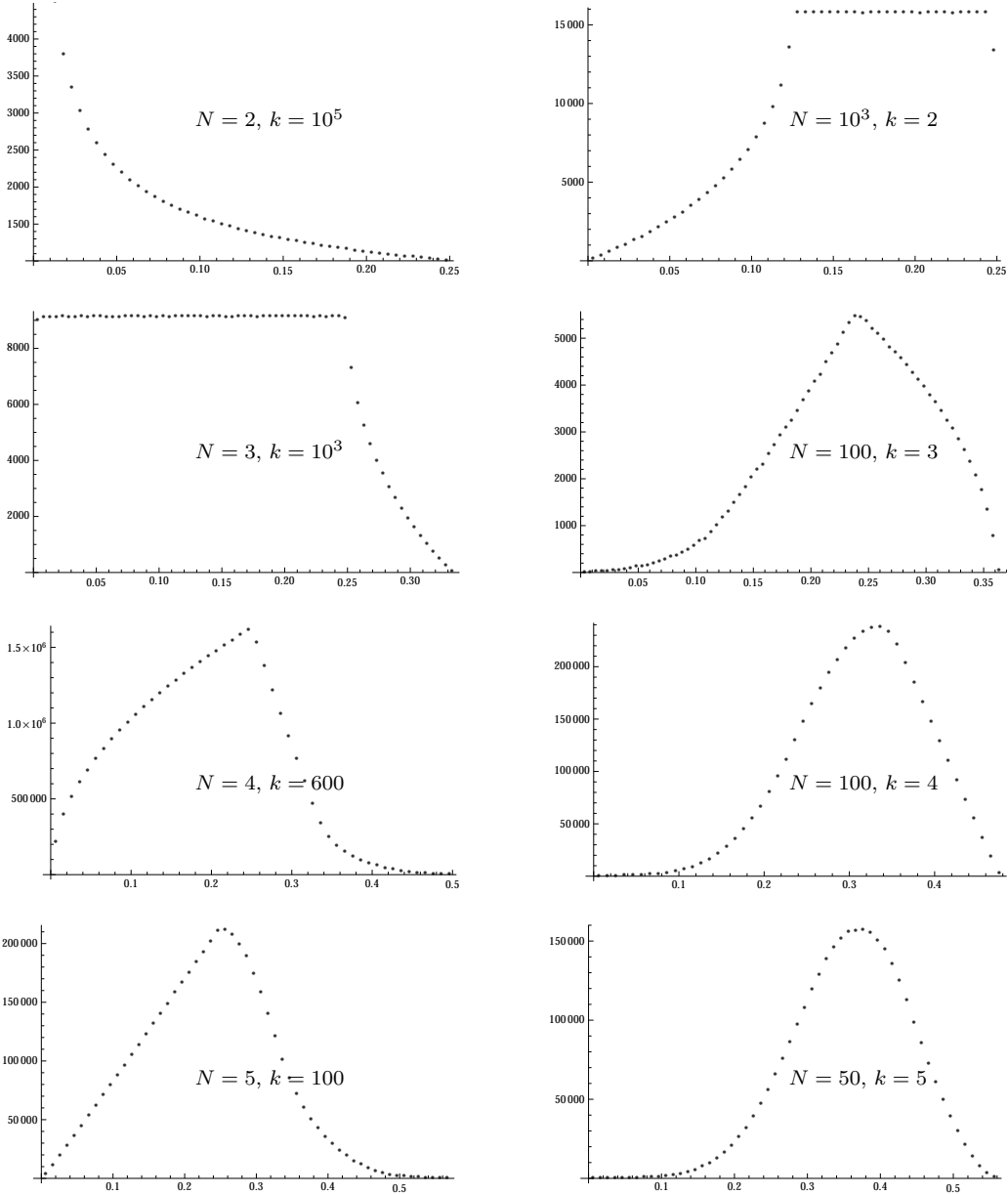

\begin{figure}[t!]
\begin{center}
\includegraphics[width=5in]{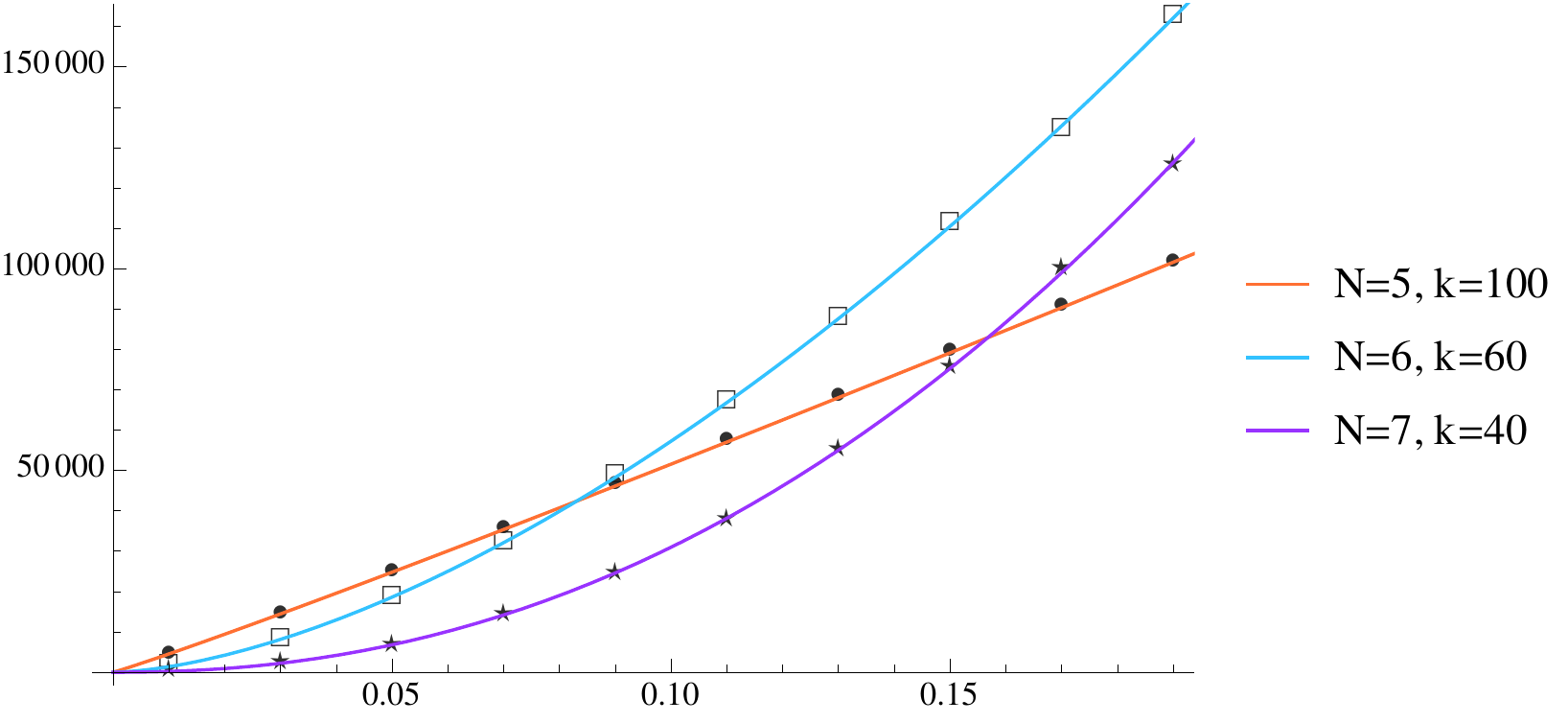} 
\begin{picture}(0,0)
\setlength{\unitlength}{1cm}
\put(-13,6){${\cal D}(\barh)$}
\put(-3,0){$\barh$}
\put(-0.2,3.48){\small{$: \alpha = 1.057$}}
\put(-0.34,2.86){\small{$: \alpha = 1.622$}}
\put(-0.34,2.28){\small{$: \alpha = 2.19$}}
\end{picture}
\caption{
The behaviour of ${\cal D}(\barh)$ for small conformal dimensions with $\lambda \approx 0$. We show the histogram data as the discrete points (differentiated by the symbol) and a polynomial fit $A\, \barh^\alpha$ for the various cases, indicating the best fit value of $\alpha$ in the legend. 
} 
\label{f:smallhlam0power}
\end{center}
\end{figure}
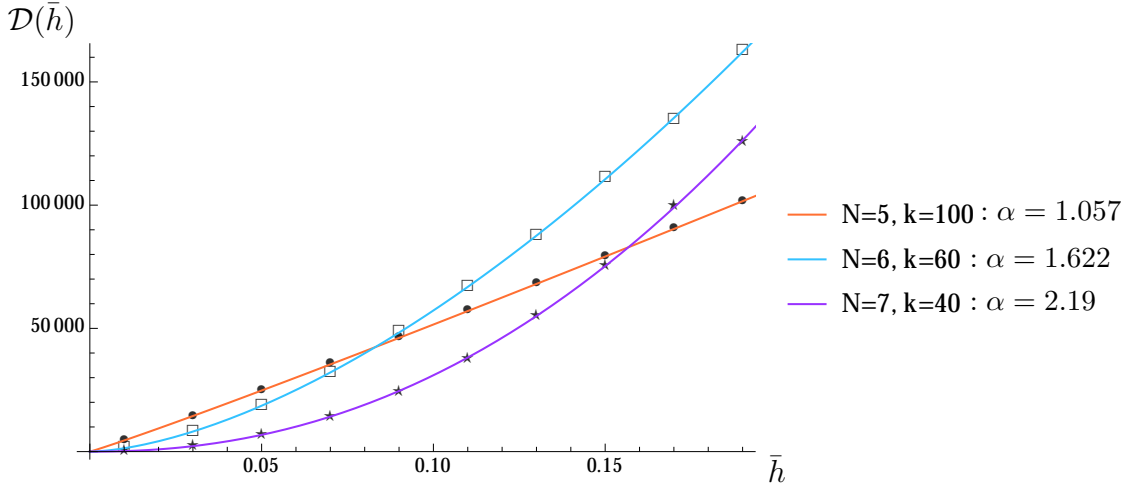

For $\lambda\approx 0$ we can describe the spectrum of light states following the analysis of 
\cite{Gaberdiel:2011aa}. As was  shown there, for $\lambda\approx 0$, by a suitable rescaling of the Dynkin labels, the light states are labelled
by $\tilde\Lambda$ with $\sum_j \tilde\Lambda_j\leq 1$, and the corresponding conformal dimension equals
\begin{equation}\label{hlightlam}
h(\Lambda,\Lambda) = \frac{1}{2} \langle \tilde\Lambda,\tilde\Lambda \rangle \ .
\end{equation}
Here $\tilde\Lambda=\frac{1}{k} \Lambda$, and the $\langle \cdot ,\cdot \rangle$ is the inner product on 
the $(N-1)$-dimensional weight space. Thus, the number of states of `length' less or equal to $r$ scales as 
$D(r)\sim r^{N-1}$, and hence 
\begin{equation}
\frac{dD}{dr} = {\rm const}\,\cdot\,  r^{N-2} \ .
\end{equation}
On the other hand, since the conformal dimension $h(r)\sim r^2$ it follows that 
\be
\frac{dD}{dh} = \frac{dD}{dr} \cdot \frac{dr}{dh} = {\rm const}\cdot r^{N-2} \cdot \frac{1}{r} = {\rm const}\cdot
 r^{N-3} = {\rm const}\cdot h^{\frac{N-3}{2}} \ .
\ee
Thus we conclude that the density of light states behaves, for $\lambda\approx 0$, as 
\be\label{poly}
{\cal D}_0(h) = C\, h^{\frac{N-3}{2}} \ . 
\ee
This description is valid up to those $h$ for which the constraint $\sum_{j} \tilde\Lambda\leq 1$ comes into play.
Since $\tilde\Lambda^0 = [\tfrac{1}{2},0,\ldots,0,\tfrac{1}{2}]$ has 
\be
h(\tilde\Lambda^0;\tilde\Lambda^0) = \tfrac{1}{4} \ ,
\ee
we conclude that (\ref{poly}) can only be trusted up to $h\leq \tfrac{1}{4}$. This is in excellent agreement with
various numerical analyses that we have done for small $N$, see Figs.~\ref{f:lam01plots} and \ref{f:smallhlam0power}. 

We should mention in passing that (\ref{poly}) actually continues to describe the {\em full} primary spectrum beyond  $h=\tfrac{1}{4}$ very well (as follows from the continuous orbifold point of view \cite{Gaberdiel:2011aa}). Indeed, the lowest non-light state seems to correspond to the representation
\begin{equation}
\Lambda^+ = \left[1;\frac{k-1}{2},0,\ldots,0,\frac{k-1}{2}\right]  , \qquad \Lambda^- = \left[0;\frac{k+1}{2},0,\ldots,0, \frac{k+1}{2}\right]
\label{lnlrep}
\end{equation}	
with conformal dimension 
\begin{equation}
h = \frac{(2N + k -1)\, (k+1)}{4\, (N+k)\, (N+k+1) } \approx \frac{1}{4}\left(1-\lambda^2\right)\ \  \to \ \ \frac{1}{4} \qquad \left(\text{when} \; k\gg N\right)  .
\label{firstnl}
\end{equation}	
Thus while for $h \leq \frac{1}{4}$ the light states are growing polynomially, their growth slows down around this point. The distribution is nevertheless monotone increasing till $\hpeak$, though at a slower rate.  The `missing' states beyond $h=\tfrac{1}{4}$ appear to be accounted for in terms of non-light primary states.

\subsubsection{The Partition Function for $\lambda\approx 0$}\label{sec:4.1.1}

To extract some more detailed physics of the spectrum, we turn from the microcanonical description adopted hitherto to a canonical one. The spectral information can be encapsulated in a partition function. In terms of the resulting free energy we can explore how the spectrum of light states influences the phase structure of the theory.  

In order to determine the form of the partition function, we first need to evaluate the 
constant $C$ in (\ref{poly}). We determine this from the normalisation condition 
\begin{equation}
\int_0^{\hmax} dh\,  {\cal D}_{0}(h) = {N+k-1 \choose k} \cong \frac{k^{N-1}}{(N-1)!} \ .
\end{equation}
In deriving the above we have used the fact that $\lambda \approx 0 \; \Longrightarrow \; k \gg N $. Furthermore, since $\hmax \cong \tfrac{N}{8}$, this leads, up to unimportant numerical coefficients (and subleading terms), to 
\begin{equation}
C \cong \frac{k^{N}}{N^{\frac{N}{2}} \, N!} \ . 
\end{equation}
Now that we have fixed the normalisation of the distribution, we can evaluate the partition function as 
\begin{equation}
Z = \int_0^{\infty} dh \, {\cal D}_{0}(h) \, e^{-4\pi \tau_2 h} = 
\frac{k^N}{N^{\frac{N}{2}} \, N!} \, \bigl( \tfrac{N}{2})!\, T^{\frac{N}{2}} \ ,
\end{equation}
where we introduce the thermal scale via $4\pi \tau_2 = T^{-1}$.\footnote{We use the conventional parameterisation in terms of the modular parameter $\tau$ of the two-torus to compute the thermal partition function of a CFT in 2 dimensions.} Thus we obtain, using Stirling's formula, and ignoring (as always) subleading terms
\begin{equation}\label{4.11}
\log Z = \frac{N}{2} \, \log \Bigl(\frac{k^2}{N^2} T\Bigr) = \frac{N}{2} \, \log \Bigl(\frac{T}{\lambda^2}\Bigr) \ ,
\end{equation}
where we have used that, for $k\gg N$, $\lambda=\tfrac{N}{N+k} \cong \tfrac{N}{k}$.  This reproduces the result derived in \cite{Banerjee:2012aj}. Later we shall see how one can directly obtain the partition function from the description in terms of free non-relativistic fermions (in a regime where $\lambda \ll T$).

\subsection{The Situation near $\lambda\approx 1$} 
\label{sec:lam1}

We can also estimate the distribution of the light states for $\lambda\approx 1$. One way to approach this problem
is to use the relation between the quadratic Casimir of representations of $\mathfrak{su}(N)_k$ and
its level-rank dual $\mathfrak{su}(k)_N$. As is explained in \cite[eq.~(1.1)]{Naculich:1990hg}, 
we have the relation 
\begin{equation}\label{C2rel}
C_2(\Lambda)^{N,k} = (N+k) \frac{B_\Lambda}{2} \Bigl( 1 - \frac{ B_\Lambda}{Nk}\Bigr) - C_2(\hat\Lambda)^{k,N} \ , 
\end{equation}
where $\hat\Lambda$ is the flipped Young diagram, where rows and columns have been interchanged, 
and $B_\Lambda$ is the number of boxes of $\Lambda$. (In the following we shall write $\hat{N}=k$ and
$\hat{k}=N$ for the rank and level of the dual description, respectively.) Furthermore, in order for 
this identity to hold we have to assume that $\Lambda$ does not have $k$ boxes in the first row, but 
only at most $k-1$ --- otherwise $\hat\Lambda$ is not an allowed representation of $\mathfrak{su}(k)$, unless
we remove the first row, etc.

We are interested in applying this formula to the case where $\lambda\approx 1$, i.e., where $N\gg k\gg 1$, for which we 
want to determine the spectrum of light states with
\begin{equation}\label{hlrrel}
h(\Lambda) = \frac{C_2(\Lambda)^{N,k}}{(N+k)(N+k+1)}  = \frac{B_\Lambda}{2(N+k+1)}  \Bigl( 1- \frac{B_\Lambda}{Nk}\Bigr) 
- \hat{h}(\hat\Lambda) \ ,
\end{equation}
where $\hat{h}(\hat{\Lambda})$ is the conformal dimension of the light state associated to the 
flipped representation in the theory with $\hat\lambda = \tfrac{\hat{N}}{\hat{N}+\hat{k}} \cong 0$.

As explained in the previous subsection, the distribution of light states for $\hat\lambda\approx 0$
is described by (\ref{poly}), 
where the limiting case with $h=\frac{1}{4}$ comes from the representation with Dynkin labels 
$[\hat{k}/2, 0, \ldots , 0, \hat{k}/2]$, which has $\hat{N} \hat{k} /2 = N k/2 $ boxes. If we are relatively close
to this case --- this is in a sense where (\ref{poly}) is best --- then the first term becomes approximately
\begin{equation}
\frac{B_\Lambda}{2(N+k+1)}  \Bigl( 1- \frac{B_\Lambda}{Nk}\Bigr)  = \frac{Nk}{8(N+k+1)} =  \frac{N}{8}(1-\lambda) = \hmax \ . 
\end{equation}
In this regime we therefore have
\begin{equation}
\hat{h}(\hat\Lambda)  =  \hmax -  h(\Lambda)  \ .
\end{equation}
Thus the distribution of $h$ near $\hmax$  is given by 
\begin{equation}
\lambda\approx 1 , \ 
|h-\hmax|\ll 1 : \qquad  {\cal D}_1( h) \sim (\hmax-h)^{\frac{\hat{N}-3}{2}} = (\hmax-h)^{\frac{k-2}{2}} \ ,
\end{equation}
where we have used in the last step that because of the subtlety described below (\ref{C2rel}) it is more
natural to identify $\hat{N}-1\equiv  k$. This prediction is very well supported by numerical results; for example 
comparing the two columns in \fig{f:lam01plots} we see that the spectrum indeed obeys the level-rank exchange 
even for small values of $N$ and $k$.\footnote{Note that we should compare the figures for which 
$k=N-1$, i.e., the relevant figure on the right is one above the one on the left.}

We can also try to estimate the distribution of states at $\lambda\approx 1$ for small conformal dimensions. At 
$\hat\lambda\approx 0$ the distribution at small conformal dimensions is well approximated by (\ref{poly}) 
since the conformal dimension is roughly of the form (see eq.~(\ref{hlightlam}))
\begin{equation}
\hat{h} = \frac{1}{2}\,  (\tilde\Lambda,\tilde\Lambda) \ , \qquad 
\tilde\Lambda = \frac{\hat{\Lambda}}{\hat{k}} \quad \hbox{with}\ 
\sum_j \tilde\Lambda_j \leq 1 \ .
\end{equation}
For these Young diagrams, the conformal dimension of the light state associated to the flipped $\Lambda$ is then
again given by (\ref{hlrrel}), i.e., as 
\begin{equation}\label{hLam}
h(\Lambda) = \frac{\tilde{r}}{2} \Bigl( 1 - \frac{\tilde{r}}{k} \Bigr) - \frac{1}{2} (\tilde\Lambda,\tilde\Lambda) \ ,
\end{equation}
where $\tilde{r}$ is the number of boxes of the rescaled representation $\tilde\Lambda$ (with
$\tilde{r}\leq k$). If we denote by $R$ the `length' of the rescaled representation $\tilde\Lambda$, i.e., 
\begin{equation}
(\tilde\Lambda,\tilde\Lambda) = R^2 \ ,  \qquad \hbox{then} \qquad
\tilde{r} \sim R \ .
\end{equation}
Thus the conformal dimension $h(\Lambda)$ is proportional to $R$ (rather than $R^2$) --- this is at least
true for sufficiently small representations. Then the density becomes
\begin{equation}
\lambda\approx 1,\  h\ll 1 : \qquad {\cal D}_{1}(h) \sim h^{\hat{N}-2}  \sim h^{k-1}  \ .
\label{lam1smh}
\end{equation}

\medskip

There is actually another way at which one may arrive at the same conclusion. We know that the $\lambda=1$ 
theory can be described in terms of free bosons. More specifically, the theory where $N\rightarrow \infty$
with $k$ fixed (i.e., with $\lambda=1$) has $c=2k$, and is described by $k$ complex bosons. However, as for the free fermion
theory at $\lambda=0$, we have to impose a singlet condition with respect to ${\rm U}(k)$ in order to 
describe the ${\cal W}_{\infty}[1]$ theory  --- otherwise,
the vacuum representation (i.e., the spectrum of purely left-moving states) would be too large. But then,
as for the free fermion theory, we need to include also the twisted sectors, and their ground states
will be the lightest states (at least for sufficiently small $h$). 

The twisted sectors are again labelled by elements in the Cartan torus of ${\rm U}(k)$, and for a given
twist $\nu\in \mathbb{R}^k$, the conformal dimension of the ground state equals
\begin{equation}
h(\nu) = \frac{|\nu|}{4}  \bigl(1-|\nu| \bigr) \ .
\end{equation}
This then has the same structure as (\ref{hLam}), remembering that 
$\tilde{\Lambda}\in\mathbb{R}^{\hat{N}-1}\cong \mathbb{R}^{k}$.
\medskip

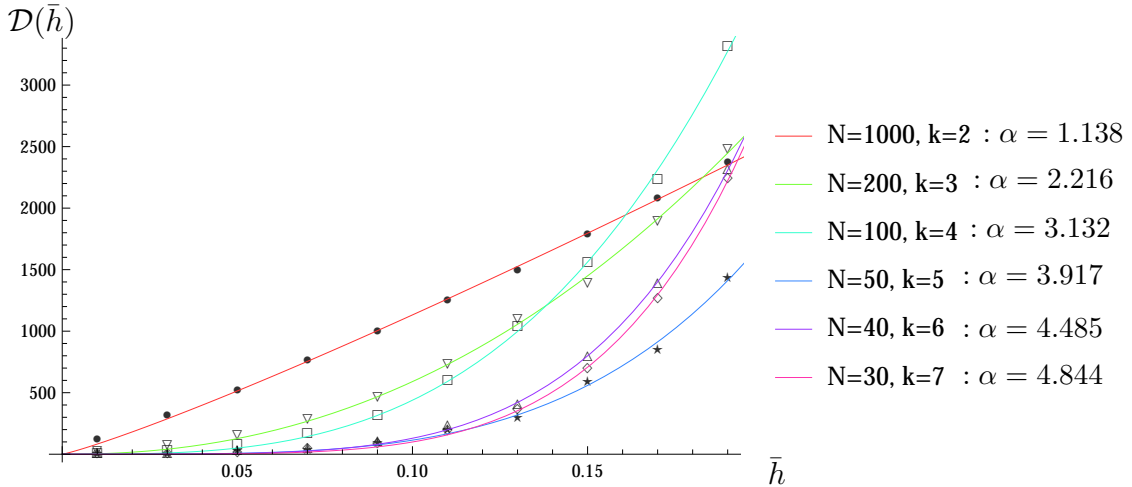
\begin{figure}[t!]
\begin{center}
\includegraphics[width=5in]{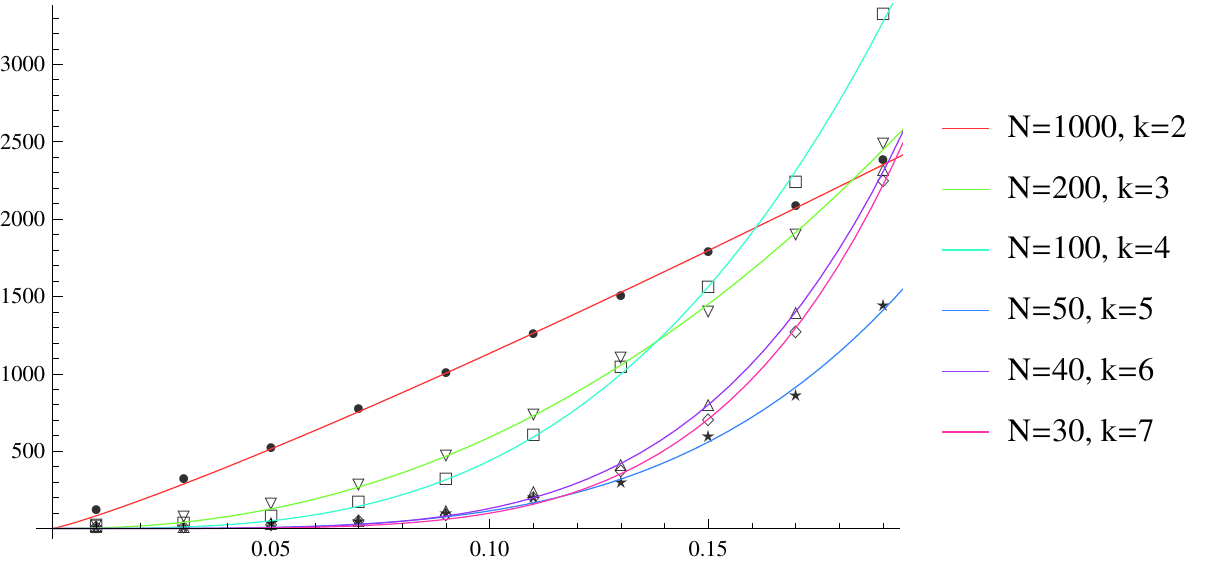} 
\begin{picture}(0,0)
\setlength{\unitlength}{1cm}
\put(-13,6){${\cal D}(\barh)$}
\put(-3,0){$\barh$}
\put(-0.17,4.53){\small{$: \alpha = 1.138$}}
\put(-0.3,3.92){\small{$: \alpha = 2.216$}}
\put(-0.34,3.27){\small{$: \alpha = 3.132$}}
\put(-0.44,2.65){\small{$: \alpha = 3.917$}}
\put(-0.44,1.95){\small{$: \alpha = 4.485$}}
\put(-0.44,1.32){\small{$: \alpha = 4.844$}}
\end{picture}
\caption{
The behaviour of ${\cal D}(\barh)$ for small conformal dimensions with $\lambda \approx 1$. We show the histogram data as the discrete points (differentiated by the symbol) and a polynomial fit $A\, \barh^\alpha$ for the various cases, indicating the best fit value of $\alpha$ in the legend. For small values of $k$ this compares very favorably with \eqref{lam1smh}; the deviation for $k=6,7$ from the prediction is because it is numerically difficult to get to larger values of $N$ to explore the asymptotic $\lambda\to 1$ regime.
} 
\label{f:smallhlam1power}
\end{center}
\end{figure}

We have also tested these predictions numerically, cf., \fig{f:smallhlam1power}.
For the behaviour near $h = 0$ we look at the explicit spectrum for $k = \{ 2,\ldots,7 \}$ with various values of $N$. While it is 
easy to generate data for $k =2$ and $N = 1000$, it becomes prohibitive to generate the spectral data for larger values of $k$. 
The best we can do for instance is $k =7, N=30$. While the data is not very cleanly amenable to a fitting analysis (due to issues 
with binning) it seems to fit the analytical predictions well. For example we find that for $h \ll 1 $
\begin{align}
k = 2, N = 10^3 : & \qquad  {\cal D}_{1}(h) \sim 15563.7 \,h^{1.1382} 
\nonumber \\
k = 3 , N = 200 : & \qquad  {\cal D}_{1}(h) \sim 97129.2 \, h^{2.2165}
\nonumber \\
k = 7 , N = 30 : & \qquad {\cal D}_{1}(h) \sim 6.92279\times 10^6\, h^{4.8441}  
 \label{5.56}
\end{align}
which supports our estimate above.

\subsubsection{The Partition Function for $\lambda\approx 1$}

We can also now redo, for $\lambda\approx 1$, the analysis of \S\ref{sec:4.1.1}. 
Now we have 
\begin{equation}
{\cal D}_{1}(h) = C \, h^{k-1} \ , 
\end{equation}
and the normalisation condition is 
\begin{equation}
\int_0^{\hmax} dh\, {\cal D}_{1}(h) = \frac{C}{k} \Bigl(\frac{k}{8}\Bigr)^k \ . 
\end{equation} 
Thus, up to subdominant terms,
\begin{equation}
C \cong  c \frac{N^k}{k^{k} k!} \ ,
\end{equation}
and the partition function becomes
\begin{equation}
Z  = \int_0^{\infty} {\cal D}_{1}(h)\, e^{-4\pi \tau_2 h} = 
c\, \frac{N^k}{k^k} \, T^k   \ ,
\end{equation}
thus leading to
\begin{equation}\label{4.27}
\log Z  = k \, \log\Bigl( \frac{T}{1-\lambda}\Bigr) \ .
\end{equation}
Here we have used that, for $N\gg k$, $1-\lambda = \tfrac{k}{N+k} \cong \tfrac{k}{N}$. 
Again, we will reproduce this result in the next section from an exact analysis (which will be seen to be valid in a regime where $(1-\lambda) \ll T$).

\section{The Free Fermion Approach}\label{sec:ff}

We now describe an analytic approach which will allow us, in principle, to determine the distribution of the light states 
and directly the contribution to the free energy from such primaries. In particular, we will be able to prove the triangular 
form of the peak distribution that was numerically seen in \S\ref{sec:peak}.
We will also be able to make various weak coupling $(\lambda \ll 1)$ and/or low temperature $(\beta \gg 1)$ 
expansions of the expression for the free energy. 

We start with the canonical partition function 
\begin{equation}
Z_N(\lambda, \beta)= \sum_{\Lambda \in P_{N,k}}e^{-\beta \, h(\Lambda)}  \ ,
\label{lightZ}
\end{equation}
where $P_{N,k}$ denotes the set of allowed representations of $\mathfrak{su}(N)_k$, and the conformal dimension is
$h(\Lambda)=\frac{\lambda^2}{N^2}C_2(\Lambda)$. As before, the possible representations are 
labelled by Young diagrams with at most $k$ columns and $(N-1)$ rows. Using the orthogonal basis introduced in \S\ref{sec:gen}, see eq.~(\ref{CasimirSq}),
we rewrite the dimension as
\begin{equation}
h(\Lambda) \equiv h (\{n_i \}) = \frac{\lambda^2}{2N^2} \Bigl[\sum_{i=1}^{N} n_i^2 - \frac{(\sum_i n_i)^2}{N} 
- \frac{N(N^2-1)}{12} \Bigr]\ ,
\label{hpart}
\end{equation}
where the $n_i$ $(i=1, \ldots ,N-1)$ are distinct integer spaced numbers ($n_N$ is fixed) 
which take values in the range specified in \eqref{niconstraint}. We can think of these as momentum labels of 
$(N-1)$ fermions. 
Thus we can write the partition function as 
\begin{equation}
Z_N(\lambda, \beta)= \sum_{\{n_i\}\text{ distinct}}e^{-\beta \,h(\{ n_i\})} \ .
\end{equation}

We would like to view the partition function computation as a problem of fermions in one dimension. However, as it stands, the $h(\{n_i \})$ is a Hamiltonian of interacting fermions because of the presence of cross terms in the middle term of  \eqref{hpart}. This complication is easily remedied by a simple trick which involves the introduction of an auxiliary variable. To this end we write the partition function as 
\begin{equation}
Z_N(\lambda, \beta)= \sum_{\{n_i\} \text{ distinct}}\sqrt{\frac{\beta N}{\pi\lambda^2}}
\int_{-\infty}^{\infty} dy\, e^{-\beta \frac{y^2N}{2\lambda^2}+\beta E_F} \,
e^{-\beta \frac{\lambda^2}{2N^2}\sum_i n_i^2 -\beta \frac{y}{N}(\sum_i n_i)} 
\label{fullpart}
\end{equation}
since doing the Gaussian integral over $y$ gives back the original partition function.
Here $E_F =\frac{\lambda^2}{2N^2} \times \frac{N(N^2-1)}{12}$ is just the constant shift in the energy (`filled Fermi sea') 
coming from the last term in \eqref{hpart}.
Next we recognise the nontrivial part of the partition function to be 
\begin{equation}\label{ZNlby}
Z_N(\lambda, \beta, y) \equiv \sum_{\{n_i\} \text{ distinct}} e^{-\beta \frac{\lambda^2}{2N^2}\sum_i n_i^2 -\beta \frac{y}{N}(\sum_i n_i)} 
=  \sum_{\{n_i\} \text{ distinct}} e^{-\beta \sum_i \epsilon_i} \ , 
\end{equation}
where 
\begin{equation}
\epsilon_i = \frac{\lambda^2}{2N^2} \, n_i^2 +\frac{y}{N} \, n_i 
\label{dispers}
\end{equation}
are the single particle energy levels. 
Thus the partition function  $Z_N(\lambda, \beta, y)$ is that of $(N-1)$ {\it free} fermions with a single particle dispersion 
relation given by \eqref{dispers}. The full partition function in \eqref{fullpart} is therefore
\begin{equation}
Z_N(\lambda, \beta)=e^{\beta E_F}\sqrt{\frac{\beta N}{\pi\lambda^2}}\int_{-\infty}^{\infty} dy \, 
e^{-\beta \frac{y^2N}{2\lambda^2}} \, Z_N(\lambda, \beta, y) \ .
\label{intpart}
\end{equation}

We can now apply standard methods from free fermion statistical mechanics to compute various quantities in the 
thermodynamic (large $N$) limit. For this, as we know from the conventional treatment of such systems, it is easier to 
go to the grand canonical ensemble
\begin{equation}
Z(\lambda, \beta, \mu)=\sum_{N=1}^{\infty}Z_N(\lambda, \beta)\, e^{\beta \mu N}\ ,
\end{equation}
where $\mu$ is the chemical potential. Correspondingly we also define 
\begin{equation}
Z(\lambda, \beta, \mu, y)=\sum_{N=1}^{\infty} Z_N(\lambda, \beta, y)\, e^{\beta \mu N} \ ,
\end{equation}
where $Z(\lambda, \beta,  y)$ was defined in (\ref{ZNlby}). In the thermodynamic limit we can then get the answer for the 
$Z_N(\lambda, \beta, y)$ by working in the grand canonical ensemble, 
and fixing $\mu$ (the saddle point in the sum over $N$) such that the expectation value of the number operator is $N$.

For taking the large $N$ limit we will find it convenient to introduce the continuum `momentum' 
$p=\frac{n}{N}$, for which the dispersion relation (\ref{dispers}) becomes in the thermodynamic limit
\begin{equation}
\epsilon(p) = \frac{\lambda^2}{2}p^2 +p \,y\ .
\label{contdispers}
\end{equation}
However, what is unusual is that the range of $p$ is restricted to 
\begin{equation}
p_{\rm min} \equiv  -\frac{1}{2}\;\; \leq \;p \;\leq \;\;\frac{1}{2}+\frac{(1-\lambda)}{\lambda} \equiv p_{\rm max} \ .
\end{equation}
In the grand canonical ensemble we know from the conventional Fermi-Dirac distribution that we will have, in the thermodynamic 
limit
\begin{equation}
\log Z(\lambda, \beta, \mu, y)  \equiv N\, Q=   N\int_{p_{\rm min}}^{p_{\rm max}} dp \ln (1+e^{-\beta(\epsilon(p)-\mu)} )
\label{fdpart}
\end{equation}
with the occupation number distribution given by
\begin{equation}
\rho(p) = \frac{1}{1+e^{\beta(\epsilon(p)-\mu)}}  \ .
\label{fdocc}
\end{equation}
Here we employ the form of $\epsilon(p)$ in the continuum, as given in \eqref{contdispers}. Then we have to 
solve for $\mu$ in terms of $\beta, y$; this is determined by the saddle point equation
\begin{equation}
\int_{p_{\rm min}}^{p_{\rm max}} dp\, \rho(p) 
= \int_{p_{\rm min}}^{p_{\rm max}} dp\,  \frac{1}{1+e^{\beta(\frac{\lambda^2}{2}\,p^2 +p\, y-\mu)}} =1\ .
\label{mudet}
\end{equation}
Putting in this value of $\mu= \mu(\lambda, \beta, y)$ determines the grand canonical free energy in \eqref{fdpart} as a function of 
$(\lambda, \beta, y)$. 

The integrals in \eqref{fdpart} and \eqref{mudet} cannot be done analytically. They can, however, 
be expressed in terms of incomplete Fermi-Dirac functions (since the upper limit in the integrals does not go to infinity). 
In the following subsections we give the results of an analysis of these expressions for the chemical potential as well as the 
grand canonical free energy in various limits. 

The free energy of the canonical ensemble is then given by the usual Legendre transform
\begin{equation}
\beta \,F_N(\lambda, \beta, y) \equiv - \ln Z_N(\lambda, \beta, y) =  -N\,Q+ N\ln{z} \ ,
\label{saddle}
\end{equation}
where $Q =\frac{1}{N} \,\log Z(\beta, \lambda, \mu, y)$ is defined in \eqref{fdpart}, and we introduce the {\em fugacity} $z=e^{\beta\mu}$ with $\mu=\mu(\lambda, \beta, y)$ determined above.
Finally, we  have to do the Gaussian integral over $y$ in \eqref{intpart}. 
In the large $N$ limit this can simply be done by solving the saddle point equation for $y$, i.e., 
\begin{equation}
\beta N y+ {\partial (\beta F_N) \over \partial y} = \beta N y -N{\partial Q \over \partial y} +
N{\partial \ln{z} \over \partial y} = 0 \ .
\label{saddle}
\end{equation}
Plugging this value of $y$ back into $Q(\lambda, \beta, y)$ and $F_N(\lambda, \beta, y)$ 
then finally gives us the canonical and grand canonical free energies  in the large $N$ limit as a function of $(\beta, \lambda)$.

\subsection{The Peak Distribution}
\label{sec:peakdist}

We can obtain the peak of the distribution by looking at  the high temperature limit ($\beta \rightarrow 0$), 
where the dominant contribution to the partition function will be from the states which contribute entropically the most. 
In this high temperature limit, as can be seen explicitly from \eqref{fdocc}, we have a uniform distribution of all fermion levels. The chemical potential $\mu$ or better the fugacity $z=e^{\beta\mu}$ is determined by \eqref{mudet} as
\begin{equation}
\int_{p_{\rm min}}^{p_{\rm max}} dp\, \rho(p) = \int_{p_{\rm min}}^{p_{\rm max}} dp\,  \frac{1}{1+z^{-1}} =1\ , 
\label{mudet2}
\end{equation}
which implies $z^{-1}= \frac{(1-\lambda)}{\lambda} =\frac{k}{N}\equiv r_{\rm max}$. 

We can translate the equilibrium distribution $\rho(p)$ to a saddle point shape for the distribution of the Young diagrams. This is because 
\begin{equation}
\rho(p)= -\frac{\partial x}{\partial p} \ ,
\label{pprofile}
\end{equation}
where $x={i \over N}$ is the continuum label for the fermions, and ${\Delta x \over \Delta p}\times \Delta p$ 
describes  the fraction of the fermions that occupy the momenta between $p$ and $p+\Delta p$ (the number of `$i$'s). 
But we know from \eqref{nidef} that we can translate $p(x)$ 
to a row distribution $r(x)$ by going to the continuum version
\begin{equation}
p(x)= r(x)+{1\over 2} -x\ .
\label{rowprofile}
\end{equation}
For the case of the uniform high temperature distribution we find, using \eqref{pprofile}, that 
$p(x)= -{1\over \lambda}x +c$, where $c$ is a constant. This is equivalent to 
$r(x)= -r_{\rm max}x - {1\over 2} +c$. We can fix $c$ by demanding that $r(1)=0$ since the $N^{\rm th}$ row has zero length 
by definition. Thus we arrive at 
\begin{equation}
r(x)= r_{\rm max}(1-x)\ , 
\label{rowans}
\end{equation}
which clearly describes a triangular profile. Since $r(0)=r_{\rm max} =\frac{k}{N}$, this implies that the 
first row is of length $k$. So it is a triangle of length $k$ and height $N$, exactly as the numerical simulations predict. 

\subsection{Free Energy when $\lambda \approx 0$ and $\lambda \approx 1$}
\label{sec:5.2}

In general, the presence of the cutoffs on the momentum makes the Fermi-Dirac distributions 
analytically complicated to tackle. However, one can in principle write down expressions in terms 
of incomplete Fermi-Dirac functions and perform systematic expansions. 
Near the end points, i.e., when $\lambda \rightarrow 0,1$, there are some further simplifications and we describe this analysis in the following. A large part of the analysis can be carried out for any temperature but at some stage we will specialise to low temperature.

\subsubsection{Fermions in the $\lambda \to 0$ limit}
Let us start with the partition function \eqref{fdpart}. At first sight, we might think we can drop the quadratic term 
in $\epsilon(p)$, see eq.~(\ref{contdispers}),
 as $\lambda \rightarrow 0$. However, we see that when we are close to the upper limit $p_{\rm max} \sim {1\over \lambda}$, then this term does contribute to order one. 
So we rescale the variable of integration $p^{\prime}=\lambda p$. We will also rescale $y^{\prime}={y\over \lambda}$. This is a good thing to do as we see from the Gaussian term in $y$ in \eqref{intpart}. 

With these rescalings the grand canonical partition function reduces to
\begin{equation}
\log Z(\lambda, \beta, \mu, y) 
={N\over \lambda}\int_{-\frac{\lambda}{2}}^{1-\frac{\lambda}{2}} dp \ln (1+z\,e^{-\beta\,(\frac{p^2}{2} +p \,y)} ) \ . 
\label{smfree}
\end{equation}
The fugacity is determined by \eqref{mudet} (the saddle point equation for the Legendre transform from the grand canonical to the canonical ensemble) which now reads 
\begin{equation}
1= {1\over \lambda}\int_{-\frac{\lambda}{2}}^{1-\frac{\lambda}{2}} dp\frac{1}{(1+z^{-1}\,e^{\beta\,(\frac{p^2}{2} +p \,y)})}\ .
\label{smmudet}
\end{equation}
We need the integral to be of order $\lambda$ for this to hold. Since the range of $p$ is finite and if we assume (self-consistently -- as well will see later) that $y$ takes a finite value as $\lambda \rightarrow 0$, then the only way this can happen is for $z^{-1}$ to be large -- of order 
${1\over \lambda}$  -- in this limit. 
So let us take
\begin{equation}
z=\frac{\lambda}{ f_{(1)}(\beta, y)} \Bigl(1+\frac{\lambda}{h(\beta, y)}\Bigr)\  ,
\label{fugeqn}
\end{equation}
keeping terms to the first non-trivial order in $\lambda$. Then we find 
\begin{eqnarray}
\lambda& = & \int_{-\frac{\lambda}{2}}^{1-\frac{\lambda}{2}} dp\, 
\frac{z\,e^{-\beta\,(\frac{p^2}{2} +p\, y)}}{(1+z\,e^{-\beta\,(\frac{p^2}{2} +p \,y)})} \\
& =&  \frac{\lambda}{ f_{(1)}(\beta, y)} \Bigl(1+\frac{\lambda}{h(\beta, y)}\Bigr)
\int_{-\frac{\lambda}{2}}^{1-\frac{\lambda}{2}} dp \, (1-z\,e^{-\beta\,(\frac{p^2}{2} +p\, y)})\, e^{-\beta\,(\frac{p^2}{2} +p \,y)} \ .
\label{smmudet2}
\end{eqnarray}
We now equate the terms on each side order by order in $\lambda$. To leading order we have
\begin{equation}
f_{(1)}(\beta, y)= \int_0^1dp \, e^{-\beta\,(\frac{p^2}{2} +p \,y)} \ ,
\label{fdef}
\end{equation}
which can be expressed in terms of the error function.
To next order we find, taking into account the term coming from the $\lambda$ dependence in the endpoints 
\begin{equation}
\frac{1}{h(\beta, y)} = \frac{f_{(2)}}{f_{(1)}^2} -\frac{1}{2f_{(1)}} \bigl(1-e^{-\beta\,({1\over 2}+y)}\bigr) \ ,
\label{heqn}
\end{equation}
where
\begin{equation}
f_{(2)}(\beta, y)= \int_0^1 dp \, e^{-2\beta\,(\frac{p^2}{2} +p \,y)} \ .
\label{f2def}
\end{equation}
Thus we have solved for the fugacity $z= z(\lambda, \beta, y)$ up to the first two orders in a small $\lambda$ expansion \eqref{fugeqn}, where the functions $f_{(1)}(\beta, y)$ and $h(\beta, y)$ are given in \eqref{fdef} and \eqref{heqn}, respectively. 

The grand canonical free energy is obtained from \eqref{smfree}; using the expression for the fugacity to the order we have computed, we find on expanding the logarithm and 
keeping track of the contribution from the endpoints that
\begin{equation}
Q= \Bigl(1+ \lambda \frac{f_{(2)}}{2\,f_{(1)}^2}\Bigr) \  .
\label{Qpotn3}
\end{equation}
Note that we have implicitly neglected terms like $\beta\lambda$ in expanding out terms in the exponent. In other words, 
we have assume that the temperature $\lambda  \ll T$. But otherwise there is no restriction on the temperature. Putting the pieces together we obtain the canonical free energy as
\begin{equation}
\beta F_N(\lambda, \beta, y) =  -NQ+ N\ln{z} = -N\Big[(1+ \lambda \frac{f_{(2)}}{2f_{(1)}^2}) -\ln{\frac{\lambda}{f_{(1)}(\beta, y)}} -\frac{\lambda}{h(\beta, y)} \Big]\ .
\label{saddle2}
\end{equation}

The leading term for small $\lambda$ is the logarithmic piece. 
We can now take a low temperature limit $\beta \rightarrow \infty$ (while still keeping $\beta\lambda \ll 1$). It is easy to see either by using the asymptotics of the error function or more simply, through rescaling variables ($p'=\beta p$) that in this limit $f_{(1)}(\beta, y) \rightarrow \frac{1}{\beta y}$ and $f_{(2)}(\beta, y) \rightarrow \frac{1}{2\beta y}$.\footnote{We also see from here that the effective expansion parameter for the fugacity in \eqref{fugeqn} is $\beta\lambda y$ which must be much less than one. This is consistent with the earlier statement that we are assuming $\lambda \ll T$.} Therefore the free energy becomes
\begin{equation}
\beta F_N(\lambda, \beta, y) = -N\Big[(1+  \frac{\lambda\beta y}{4}) -\ln{(\lambda\beta y)}  \Big] \ .
\label{saddle3}
\end{equation}
Now we can perform the last step of doing the $y$ integral in \eqref{intpart}, remembering that we have rescaled $y$ 
by a factor of $\lambda$,
\begin{align}
Z_N\, (\lambda,\beta) &= e^{\beta\, E_F} \, \sqrt{\frac{N\,\beta}{\pi}} \, \int_{-\infty}^{\infty} \, dy' \; 
e^{-\frac{1}{2}\,N\, \beta \, {y'}^2 +N\, - N\, \log \left(\beta\, \lambda\, y'\right)}
\nonumber \\
& = \left[\frac{1}{\sqrt{\pi}}\; e^{\beta\, E_F+N} \, \left(\frac{N}{2}\right)^{\frac{N}{2}}\, \Gamma\left(\frac{1-N}{2}\right) \right] \times \left(\frac{1}{\beta\, \lambda^2}\right)^{\frac{N}{2}}
\nonumber \\
\qquad \Longrightarrow \qquad  \log Z_N  &= \frac{N}{2}\, \log \frac{T}{\lambda^2} \ .
\label{}
\end{align}	
We have kept here only the leading log dependence on $\lambda$. The subleading linear term in $\lambda$ 
in \eqref{saddle3} only contributes additional terms for the free energy that are polynomial in $\lambda$. 
Note that the argument of the logarithm is large since $T \gg \lambda \gg \lambda^2$. 

We thus see that the free fermion description reproduces the low energy density of primaries, which we earlier estimated numerically and gave an argument for in \S\ref{sec:lam0}, see eq.~(\ref{4.11}). 
The result for the partition function is of course also consistent with the earlier derivation of \cite{Banerjee:2012aj}.

\subsubsection{Fermions in the $\lambda \to 1$ limit}

Another interesting regime to consider is the one where $(1-\lambda) \ll 1$. The unusual feature of this limit is that the upper cutoff on the momentum,
\begin{equation}
p_{\rm max} \approx \frac{1}{2} +(1-\lambda) = p_{\rm min}+1 + (1-\lambda)
\end{equation}
starts to approach the naive Fermi surface. Since the ground state configuration (for any $\lambda$) is one where the fermions occupy momenta in an interval of length one, we see that for $\lambda \rightarrow 1$, we are squeezing the phase space that the fermions can occupy to the least possible one. In other words, the momentum distribution for any temperature is almost the same as the zero temperature distribution. Thus we can make a systematic approximation scheme by starting with $\rho(p) \approx 1$.  

The equation determining the fugacity is now 
\begin{align}
1
 &= \int_{-{1\over 2}}^{p_{\rm max}} dp\,  \frac{1}{1+z^{-1}\,e^{\beta\,(\frac{\lambda^2}{2}p^2 +p\, y)}} 
 \approx \int_{-{1\over 2}}^{p_{\rm max}} dp\, (1- z^{-1}\, e^{\beta\,(\frac{\lambda^2}{2}p^2 +p \,y)}) \nonumber \\
&= 1+(1-\lambda) -z^{-1}\; \tilde{f}_{(1)}(\beta, y) \ ,
\label{mudet3}
\end{align}
where     
\begin{equation}
\tilde{f}_{(1)}(\beta, y)= \int_{-\frac{1}{2}}^{\frac{1}{2}}dp \; e^{\beta\,(\frac{p^2}{2} +p \,y)} \ .
\label{fdef2}
\end{equation}
Thus to leading order in $(1-\lambda)$ we have 
\begin{equation}
z^{-1}=\frac{(1-\lambda)}{\tilde{f}_{(1)}(\beta, y)}\ .
\label{fugeqn2}
\end{equation}
It is clear that this is the first term in a systematic expansion in powers of $(1-\lambda)$. 
Once again, the free energy can be computed from \eqref{fdpart} where
\begin{align}
Q &= \int_{p_{\rm min}}^{p_{\rm max}} dp \, \ln (1+z\,e^{-\beta\,(\frac{p^2}{2} +p\, y)} ) \nonumber \\
&=  \int_{p_{\rm min}}^{p_{\rm max}} dp\,  \big[\ln\left(z\,e^{-\beta\,\epsilon(p)}\right)+z^{-1}\,e^{\beta\,\epsilon(p)} \big] \nonumber \\
&= -\left[1+(1-\lambda)\right]\ln{\frac{(1-\lambda)}{ \tilde{f}_{(1)}(\beta, y)}} 
- \frac{\beta}{6}(p_{\rm max}^3-p_{\rm min}^3) - \frac{\beta}{2}(p_{\rm max}^2-p_{\rm min}^2)y  +(1-\lambda)  \ .
\label{Qpot}
\end{align}	
Thus the full free energy is, to this order,
\begin{align}
F_N(\lambda, \beta, y)  &= -NQ+ N\ln{z} \nonumber \\
&= N\Big[(1-\lambda)\ln{\Big(\frac{(1-\lambda)}{ \tilde{f}_{(1)}(\beta, y)}\Big)} 
+ \frac{\beta}{6}(p_{\rm max}^3-p_{\rm min}^3) +\frac{\beta}{2}(p_{\rm max}^2-p_{\rm min}^2)y \nonumber \\
& \qquad \quad - (1-\lambda) \Big]\ .
\label{Qpot}
\end{align}	
We can take a low temperature limit of this expression, i.e., 
we can consider the limit where $(1-\lambda) \ll T \ll 1$. Then we can evaluate
\begin{equation}
\tilde{f}_{(1)}(\beta, y) \rightarrow \frac{1}{\beta (|y|+\frac{1}{2})}\, e^{\frac{\beta}{2}(|y|+\frac{1}{4})} \ .
\label{tilflim}
\end{equation}
Thus, apart from various constant pieces, the relevant logarithmic part of the free energy in \eqref{Qpot} is
\begin{equation}
F_N(\lambda, \beta, y) \rightarrow k \ln{\Big(\beta(y+\frac{1}{2})(1-\lambda)\Big)} + {\rm finite}.
\label{Fleqone}
\end{equation}
This is not modified as we perform the saddle point integration over $y$, i.e.,  we continue to obtain as the leading piece
\begin{equation}
F_N(\lambda , \beta) \approx - k \ln{\frac{T}{(1-\lambda)}} +{\rm  finite}.
\label{Fleqone2}
\end{equation}
The logarithmic behaviour of the free energy is indeed as predicted by the analysis of the low-lying light primary states, see 
eq.~(\ref{4.27}) in \S\ref{sec:lam1}.

\subsection{Free energy for generic $\lambda$}
\label{sec:lamoth}

We have seen above that the free fermion picture corroborates the results of the numerical investigations and provides evidence for the  picture developed in \S\ref{sec:lam01}. 
Thus $\lambda \rightarrow 0,1$ are the only regimes where there appears to be a non analytic dependence of the free energy on $\lambda$. Nevertheless there seems to be no phase transition as a function of $T$. For generic values of $\lambda$, as one might already anticipate from the numerical experiments of \S\ref{sec:shape}, there is no interesting feature.
We should be able to ascertain this from the statistical mechanics of the fermions we have been discussing. This is indeed possible, albeit a bit involved, owing to the limits on the momenta. We explain how to adapt the standard Sommerfeld expansion for analysing the low temperature behaviour of
fermions in Appendix~\ref{sec:lowT} to obtain a result for the free energy at low temperatures $T \ll 1$. While much of the analysis described there is valid for any $\lambda \in (0,1)$, it should be noted that we work in a complementary regime $T \ll \{\lambda, 1-\lambda\}$ to the one we have used above. In any event we see no signs of any non-analytic behaviour at low temperatures, consistent with the general expectation.

\section{The Rest of the Spectrum}
\label{sec:restspec}

\paragraph{The spectrum of non-light states:}
The main focus of this paper has been the spectrum of light states encountered in the $W_{N,k}$ minimal models in the 't Hooft limit. As we have discussed earlier, these are but a small subset of the entire primary spectrum, their main distinguishing feature being that there are representatives of such states whose conformal dimension is vanishingly small. However, these light states span over a range
$[0,\hmax]$ with $\hmax \sim N$ as described in \S{\ref{s:3.1}}. So at some point we would have to confront the fact that there are other primaries in the spectrum. We now turn to some of the key features of the spectrum of non-light states.

The main thing to note about the non-light states is that there are an exponentially larger number of them. Since these states are classified by two distinct representations $(\Lambda^+;\Lambda^-)$ we infer that the number of such states is given by ${\cal N}_{N,k}^2/N = e^{2\,N\,G(\lambda)}/N$, with the factor of $N$ accounting for the ${\mathbb Z}_N$ automorphism symmetry 
of eq.~(\ref{fieldid}), and $G(\lambda)$ given in \eqref{Glam}. This fact alone makes them hard to work with numerically, since we are forced to rely on data generated for small values of $N$ and $k$. Nevertheless, we have managed from our numerical experiments to glean some basic facts about these states, which fits well with certain heuristic arguments.
\begin{itemize}
\item The non-light primaries enter the spectrum at $h = \frac{1}{4} \left(1-\lambda^2\right)$ as described in \S\ref{sec:lam0}, cf., eq.~\eqref{firstnl}. Note that at this point the light state part of the spectrum is still monotonically growing and we have relatively few non-light states.
\item The maximum conformal dimension attained in the spectrum is ${\cal O}(N^3)$. In fact, we find
the dimension is maximised when each of the individual representations has a single non-zero entry which are maximally separated from each other (accounting for the cyclic symmetry). To wit,
\begin{align}
\hmax^{\rm f} &= \frac{1}{8} \left(\frac{1-\lambda}{\lambda}\right)^2 N^3\,, 
\nonumber \\
 \Lambda^+_j = k\, \delta_{j,j_+}\,,\quad &\Lambda^-_j = (k+1)\, \delta_{j,j_-}\,,\qquad |j_+-j_-| = \frac{N}{2} \ .
\label{genmax}
\end{align}	
\item The spectrum has a characteristic peak at $h \sim {\cal O}(N^2)$. Numerical investigations show that 
\begin{equation}
\hpeak^{\rm f} = \frac{1}{24} \, \frac{1-\lambda}{\lambda^2}\; N^2 \,.
\label{genpeak}
\end{equation}	
In fact, from our numerical experiments we see that the density of states is always monotone increasing  in $[0,\hpeak^{\rm f}]$. 
The states near the peak are once again dominated by approximately triangular Young tableaux. Curiously, the fluctuations of the rows about the triangular representation follows a semi-circular distribution $\frac{1}{2}\left(\delta r_i^+ + \delta r_i^-\right) \approx \sqrt{ \frac{1-\lambda}{2\, \lambda^2}}\, \sqrt{\frac{i\,(N-i)}{N}}$. 

It would be nice to understand the bulk interpretation of these (very abundant) primaries of dimension $N^2$ as well as those which scale as $N^3$. In some ways they are reminiscent of the exotic branes that have been investigated recently \cite{deBoer:2012ma} which are much heavier than the conventional soliton solutions of string theory that behave as $\frac{1}{g_s}$ or $\frac{1}{g^2_s}$. 
\item The shape of the distribution can be obtained from numerical studies to be of the form of a power law modulated by an exponential (the so called Gamma-distribution). A very crude fit suggests a beguilingly simple form
\begin{equation}
{\cal P}_{\rm gen}(h) = \frac{4\,h^2}{(\hpeak^{\rm f})^3}\, \exp\left(-2\, \frac{h}{\hpeak^{\rm f}}\right) \ .
\label{gamdist}
\end{equation}	
\end{itemize}

\begin{figure}[h]
\begin{center}
\setbox1=\hbox{\includegraphics[width=3.5in]{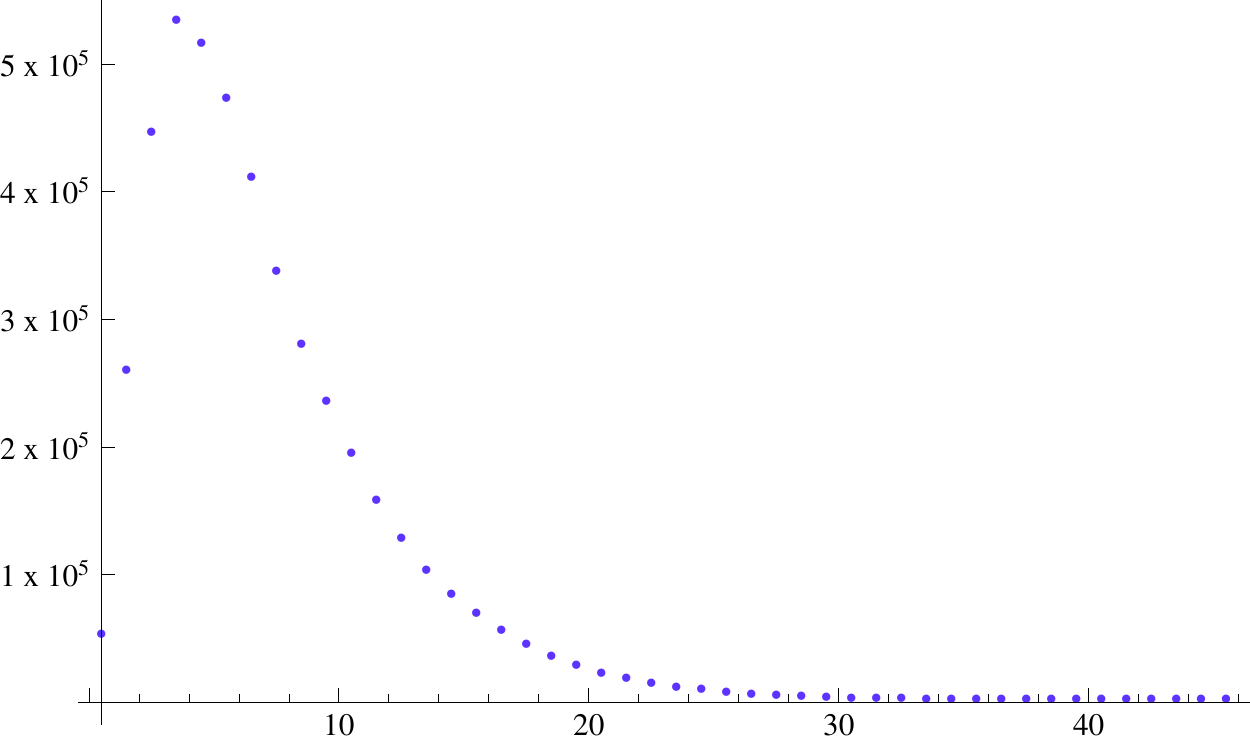}} 
\includegraphics[width=3.5in]{histogramG-N12k5}\llap{\raisebox{3cm}{
\includegraphics[width=1.6in]{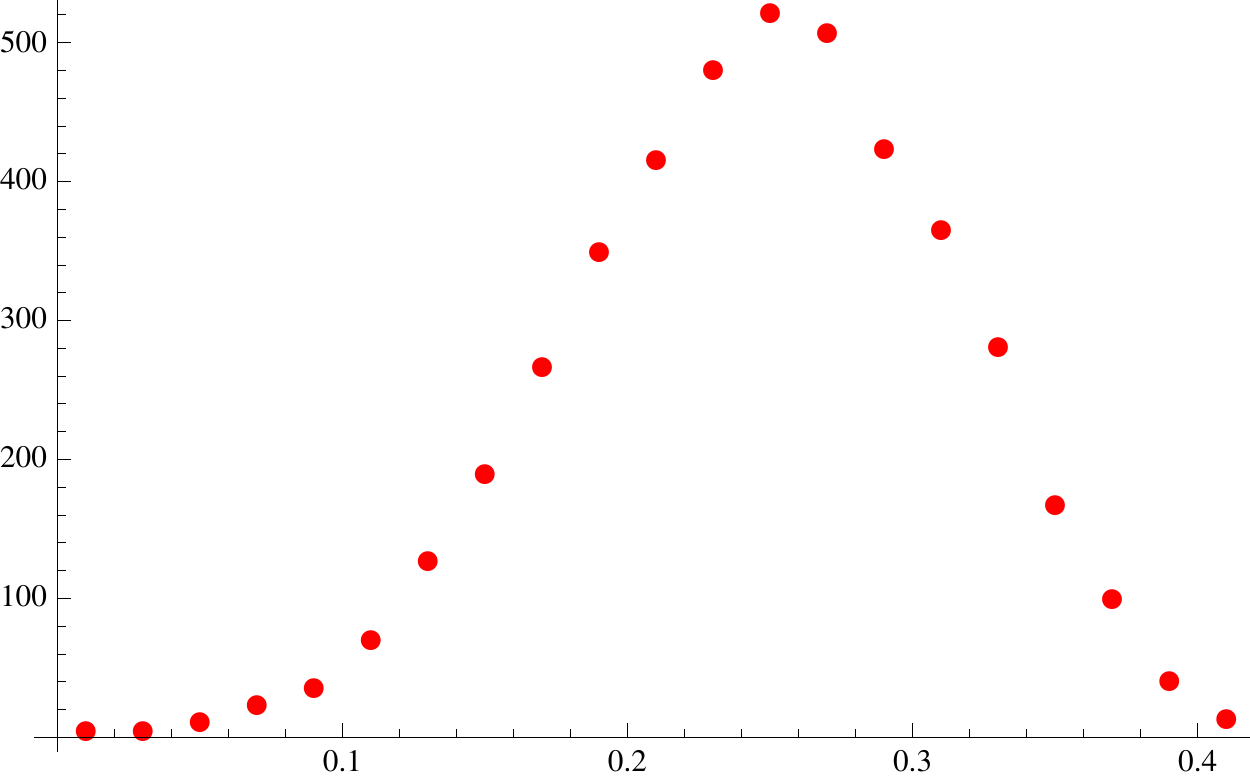}
\begin{picture}(0,0)
\setlength{\unitlength}{1cm}
\put (-5.2,2.3) {{\tiny \#\,\text{states}}}
\put (0,0) {{\tiny $h$}}
\end{picture}
}} 
\begin{picture}(0,0)
\setlength{\unitlength}{1cm}
\put (-10.7,4.5) {{\small \#\,\text{states}}}
\put (0,0) { $h$}
\end{picture}
\caption{Distribution of the conformal dimensions in the full spectrum of states, illustrated here for $N=12, k=5$. There are 
approximately $4.5 \times 10^6$ states accounting for the automorphism symmetry, while the number of light states is 
just $4368$. We also display the light state distribution in the inset for comparison. The location of the maximum 
$\hmax^{\rm f} = 45$ and the 
peak $\hpeak^{\rm f}=3.6$ are in reasonable agreement with the predictions \eqref{genmax} and \eqref{genpeak}, 
respectively (we predict $\hmax^{\rm f} = 37.5$ and $\hpeak^{\rm f} = 3.542$). Furthermore, the profile of the 
distribution is of the gamma-distribution form \eqref{gamdist}.
} 
\label{f:gendistributions}
\end{center}
\end{figure}

The key features of the general distribution are illustrated in \fig{f:gendistributions}, where we see close agreement with the analytic predictions even for relatively small values of $N$ and $k$.

Note that since the light states deplete in number following $h=\hpeak$ (see \eqref{peak}), it follows that the non-light states should start growing rapidly in the neighbourhood of $h=\hpeak$ to ensure that that total spectrum is monotone in $[0,\hpeak^{\rm f}]$. One can indeed estimate the number of non-light states to be at least of the same order as the number of light states around their peak. This follows by starting with a triangular Young tableaux for $\Lambda^+$ so that $(\Lambda^+;\Lambda^+)$ is a typical light state. One can construct a suitable $\Lambda^-$ by augmenting $\Lambda^+$ by a unit $N$-dimensional vector. The pair $(\Lambda^+;\Lambda^-)$ thus constructed accounts for the same number of states as the light states in the vicinity of $\hpeak$ (taking into account the automorphism) and satisfies the constraint \eqref{lampmk}. For $\Lambda^- = \Lambda^+ \pm \delta_{j m}$  the corresponding conformal dimension equals
\begin{align}
h(\Lambda^+; \Lambda^-) &= h(\Lambda^+,\Lambda^+) \mp \frac{1}{p}\left[\frac{m\,(N-m)}{2} + m\, r^+_{m+1} + \left(1-\frac{m}{N}\right) B^+ - B^+_{>m} \right] \nonumber \\[2pt]
& \quad + \frac{m\,(N-m)}{2\,N}\ ,
\label{}
\end{align}
where $r^+_m$, $B^+$ and $B^+_{>m}$ are the number of rows in the $m^{\rm th}$ row, the total number of boxes, and the number of 
boxes in rows greater than $m$ of $\Lambda^+$, respectively. The correction to the light state dimension is at most of ${\cal O}(N)$,
and one can verify that most of the states thus generated in fact have $h(\Lambda^+;\Lambda^-) \simeq \hpeak$.

\paragraph{Estimating the growth of descendants:}
While the general features of the spectrum of states is interesting, for the physical question of whether the theory undergoes a phase transition at some ${\cal O}(1)$ temperature, one would like to know more detailed aspects of the spectrum. For one, we not only have to worry about $W_N$ primaries at a given conformal dimension, but also at sufficiently large values of $h$ consider the contribution of $W_N$ descendants of the low-lying states. This is an even trickier proposition since different primaries have different null states; this implies that the growth of descendants can change quite dramatically depending on how we estimate the descendant contribution. 

One can  make progress by bounding the growth of descendants between two extremes. On the one hand, we can assume that there are no null states and since there are $(N-1)$ current generators $W^{(s)}$, we get a contribution
\begin{equation}
\tilde{\eta}(q)^{-(N-1)} \equiv \prod_{n=1}^{\infty}\frac{1}{(1-q^n)^{N-1}} = \sum_{m=1}^\infty \, d_{\eta,m}\, q^m\ .
\end{equation}	
On the other hand, a lower bound (at least in the 't Hooft limit) is provided by the descendants of the 
vacuum primary which is given by the modified MacMahon function
\begin{equation}
\widetilde{M}(q) \equiv \prod_{s=2}^N\prod_{n=s}^{\infty}\frac{1}{(1-q^n)} = P_N(q)\; \tilde{\eta}(q)^{-(N-1)} 
= \sum_{m=1}^\infty\, d_{W,m} \, q^m \ ,
\label{MtildeN}
\end{equation}	
where $P_N(q)= \prod_{n=1}^{N-1}(1-q^n)^{N-n}$ is a polynomial.  

The asymptotic growth of both of these can be estimated by standard techniques despite not being in the Cardy regime generically (adapting for instance results from \cite{Lucietti:2008cv}). One finds
\begin{align}
\log d_{\eta,m}&=\frac{1}{12} \left[ \sqrt{3\,N\, \left(8 \pi^2\, m+3\, N\right)} \nonumber  \right.\\
& \qquad \left.
+N \left(\frac{24 \pi ^2 \,m}{\sqrt{3\, N \left(8 \pi ^2\, m+3\,N\right)}-3 \,N}-3-6 \log (2 \pi )\right) \right.
\nonumber \\
& \left.\qquad \qquad +\;
6 \,N\,\log \left(\frac{\sqrt{3\, N \left(8 \pi ^2 m+3 \,N\right)}-3 \,N}{12 m}\right)\right]
\nonumber \\[4pt]
\log d_{W,m} &= 3\,\left(\frac{\zeta(3)}{4}\right)^\frac{1}{3} \, m^{\frac{2}{3}} + \frac{\zeta(-1)}{3}\, \log m + \left(\zeta'(-1) - \frac{\zeta(-1)}{3} \, \log (2\,\zeta(3))\ \right) \ .
\label{}
\end{align}
While these are complicated expressions, of interest to us is the growth of descendants for $m \sim N$ and $m\sim N^2$ respectively. 

\paragraph{Primaries versus descendants:}
Consider first states with $h\sim N^2$. From the asymptotics we learn that the growth of descendants is super-exponential: the Mac\-Mahon estimate  gives $\log d_{W,N^2} \propto N^\frac{4}{3}$, while  the naive eta-function estimate leads to $\log d_{\eta, N^2} \propto  N^\frac{3}{2}$. Either of these swamps the primaries, which at most grow exponentially in $N$ since the total number of primaries does no better than that. So we are led to concluding that, at these conformal dimensions, the descendants dominate the spectrum despite most of the primaries being found with these conformal dimensions.

On the other hand things are a lot more interesting when $h \sim N$. Here the MacMahon estimate leads to $\log d_{W,N} \propto N^\frac{2}{3}$, whereas the simpler eta-function estimate leads to $\log d_{\eta,N} \sim 1.722 N$. To proceed we also need to know the estimate for the number of primaries with $h\sim N$. We know from the light states having typically such conformal dimensions that there must be roughly $e^{\alpha \, N}$ such primaries with $\alpha \approx G(\lambda)$. We would ideally like to get a better estimate to ascertain if the growth is rapid enough to offset $d_{\eta,N} \sim e^{1.722 \, N}$ descendants. While we believe this is likely 
--- indeed, $G(\lambda)>1.722$ for $\lambda < 0.388$ ---  it is quite hard to get a handle on these states. These primaries are atypical in the full spectrum making it hard to accesses their behaviour for large enough values of $N$ and $k$. We nevertheless believe that primaries dominate over the descendants for $h\sim N$. It would be interesting to ascertain whether this statement implies something interesting about the dual holographic theory in the bulk: for instance is there a novel spectrum of black hole states with $h \sim N \propto c$?

\section{Discussion}
\label{sec:discuss}

Our analysis of the light states was an attempt to detect a signature of a phase transition at large $N$ in the CFT partition function. As described above, both the analytic (free fermion) and numerical approaches seem to show no sign of any change in saddle points as we change the temperature (which we take to scale as ${\cal O}(1)$). Indeed the density of states seems to be smoothly growing, and hence smoothens out any transition as conjectured in 
\cite{Banerjee:2012aj}. 

More work nevertheless needs to be done to put this conclusion on a firmer footing. 
Firstly, there is the issue of the growth of descendants which we have described in the previous section. While it seems plausible to us that descendants won't change the conclusion, this needs to be argued more carefully. Secondly, there is the issue of the non-light states which already at $h \sim N$ start dominating the spectrum over the light states. 
Again, while our numerical studies do not reveal much feature in the distribution of states, it is a bit difficult to extract this conclusion 
 reliably since this is part of the growing tail of the distribution (which peaks at $h\sim N^2$). It would therefore be desirable to have more analytic control over the full spectrum of primaries analogous to the free fermion picture for the light primaries. 

One possible approach is the following. We consider the partition function for the general primaries
\begin{equation}
Z_N(\beta, \mu^{\pm}) =\sum_{\Lambda^{\pm}}e^{-\beta ({1\over 2} (\Lambda^+-\Lambda^-)^2-\mu^
+\sum_i \Lambda^+_i -\mu^-\sum_i \Lambda^-_i)}\ .
\label{genprimpart}
\end{equation}	
This can be viewed as a grand canonical ensemble in which we have introduced chemical potentials $\mu^{\pm}$ in lieu of the constraints on the Dynkin labels $\sum_i \Lambda^{\pm}_i \leq k$ (we neglect the difference between $k$ and $k+1$ in the large $k$ limit).
We can rewrite this partition function in terms of new variables $X_i$ via
\begin{equation}
Z_N(\beta, \mu^{\pm}) =C_N\sum_{\Lambda^{\pm}}\int_{-\infty}^{\infty}\prod_{i=1}^{N-1}dX_i\,
e^{\frac{1}{2\beta}\sum_{i,j}A_{ij}X_iX_j - \sum_iX_i(\Lambda^+_i-\Lambda^-_i)
-\beta(\mu^+\sum_i \Lambda^+_i + \mu^-\sum_i \Lambda^-_i)}\ .
\label{parttransf}
\end{equation}	
Here $A_{ij}$ is the Cartan matrix of $\mathfrak{su}(N)$,
and $C_N$ is a constant coming from the square root of the determinant of this matrix and other factors irrelevant to the large $N$ limit. 
This is a useful form to write the partition function since the Cartan matrix is a discrete version of the second derivative 
\begin{equation}
\sum_{i,j}A_{ij}X_iX_j = \sum_i X_i(2X_i -X_{i-1}-X_{i+1})\ . 
\label{cartlimit}
\end{equation}	
Thus we have an integral which is `local' in the $i$ indices and hence more amenable to treating the large $N$ limit 
as a continuum limit of a local lattice model. We postpone further study of this form of the partition function to the future.

\acknowledgments 
It is a pleasure to thank Shamik Banerjee, Alejandra Castro,  Simeon Hellerman, Arnaud Lepage-Jutier, 
Alex Maloney and Eric Perlmutter for useful discussions, and Sameer Murthy for initial collaboration.
We would like to acknowledge the hospitality of the 
IPMU (Tokyo), GGI (Florence), Simons Center (Stony Brook), 
Newton Institute (Cambridge), ITF Univ.\ of Amsterdam, Stanford University, UC Berkeley, Univ.\ of Michigan (Ann Arbor), Nordita (Stockhom), YITP (Kyoto), ISM 2012 (Puri), Univ.\ of Crete (Heraklion), ICTP (Trieste), GR20 (Warsaw), Benasque Center for Science, Bogazici University (Istanbul), Jagiellonian University (Krakow), The Czech Academy of Sciences (Prague), 
Univ.\ of KwaZulu Natal (Durban), Univ.\ of Cape Town and Univ.\ of Witwatersrand (Johannesburg) during the course of this project.

The research of MRG is partially supported by a grant from the Swiss National Science Foundation. 
RG was partially supported by the Swarna Jayanthi Fellowship of the DST and more generally by the 
people of India's commitment to the fundamental sciences. 
MR supported in part by the the STFC Consolidated Grant ST/J000426/1.

\appendix
\section{Free fermions at low temperature}
\label{sec:lowT}

The free fermion model developed in \S\ref{sec:ff} can be analysed in detail using standard statistical mechanics machinery. One can explicitly derive the expressions for the free energy in terms of incomplete Fermi-Dirac integrals. However, to obtain the main physics of interest it suffices to focus attention on the low temperature behaviour of the system. As a result we will now describe a systematic approach to developing a low temperature expansion. This can be carried out for any value of $\lambda \in (0,1)$. Note that when we talk of a low temperature expansion we have in mind a temperature which is ${\cal O}(1)$ with respect to the $N$ scaling. Hence, this temperature regime probes light states whose energies still scale as $N$ while perhaps being much smaller than the peak value. 

To carry out the analysis we will find it convenient to define new variables 
\begin{equation}
\xi \equiv \frac{y}{\lambda^2} \,,  \qquad {\hat \beta} = \beta\, \lambda^2 \,, \qquad
{\bar z} = e^{\beta \mu + \frac{1}{2}\, {\hat \beta}\, \xi^2 } \,, \qquad \tau = p + \xi  \ , 
\label{newvar}
\end{equation}	
so that the single particle energy takes a simple form. Furthermore, introducing new labels for the lower and upper momentum cutoffs as $\tau_{L,R}$ as well as the zero and finite temperature Fermi momenta by $\tau_F, \tau_*$, i.e.,
\begin{equation}
\tau_L = \xi - \frac{1}{2} < \tau_F= \tau_L+1 =  \xi + \frac{1}{2} \leq \tau_R =  \xi - \frac{1}{2} +\frac{1}{\lambda} \ , \qquad \, \frac{1}{2}\,\hat{\beta}\,\tau_*^2=\ln{\bar{z}_*}\ ,
\label{taudef}
\end{equation}	
we can present the main equations for the model in a compact form.

For instance the equation determining the chemical potential \eqref{mudet} now reads 
\begin{equation}
\int_{\xi -\frac{1}{2}}^{\xi-\frac{1}{2}+ \frac{1}{\lambda}}\,  \frac{d\tau}{1+ {\bar z}_*^{-1}\, e^{\frac{1}{2}\,{\hat \beta}\, \tau^2}} = \int_{\tau_L}^{\tau_R}\,  \frac{d\tau}{1+ e^{\frac{1}{2}\,{\hat \beta}\, (\tau^2-\tau_*^2)}}  = 1  \ . 
\label{zdet}
\end{equation}	
Once we solve for ${\bar z}_*={\bar z}_*(\lambda, {\hat \beta}, \xi)$ from this equation we can compute any thermodynamic variable of interest. Using integration by parts the grand canonical free energy, see eq.~(\ref{fdpart}), 
can be decomposed as $Q=Q_0+Q_1$, where 
\begin{align}
Q_0 &= \tau_R\log\left(1 + e^{-\frac{1}{2}\, {\hat \beta}\,\left(\tau_*^2-\tau_R^2 \right)} \right)-  \tau_L \log\left(1 + e^{-\frac{1}{2}\, {\hat \beta}\,\left(\tau_*^2-\tau_L^2\right)}\right)
\nonumber \\
Q_1 &= \int_{\tau_L}^{\tau_R}\, d\tau \; \frac{{\hat \beta}\, \tau^2}{1+ e^{\frac{1}{2}\,{\hat \beta}\, (\tau^2 -\tau_*^2)}}\  .
\label{Qpot2}
\end{align}	
The canonical free energy $F_N(\lambda, {\hat \beta}, \xi)$, which is what we are ultimately interested in, is simply expressed as (rescaling the temperature)
\begin{equation}
{\hat \beta}\,F_N(\lambda, {\hat \beta}, \xi)= -N\, Q + N\,\log z_* = -N\,Q + N\, \log {\bar z}_* - \frac{1}{2}\,N\,{\hat \beta}\,\xi^2  .
\label{Znxi}
\end{equation}	
All this was for fixed $\xi$. In the end we have to do the $\xi$ (i.e., $y$) integral which amounts to solving the saddle point equations \eqref{saddle}.

To carry out the low temperature expansion, we have to identify the ground state. From 
\eqref{zdet} we see that we can think in terms of free fermions with a dispersion relation $E(\tau) = \frac{1}{2}\tau^2$. Normally, at zero temperature such fermions would occupy the states with momenta in the interval $[-\frac{1}{2}, \frac{1}{2}]$. 
However, due to the presence of the upper and lower cutoffs in \eqref{zdet} this cannot always be achieved. The precise interval will depend on the values of $\xi$ (for $\lambda$ in the interval $(0, 1)$). We find that there are five qualitatively different cases:
\begin{enumerate}
\item[(i).] $\xi < \frac{1}{2}-\frac{1}{\lambda}$. Then the filled momentum interval is $[\xi - \frac{3}{2} +\frac{1}{\lambda},  \xi - \frac{1}{2} +\frac{1}{\lambda}]$. This is purely on the negative axis and strictly above the lower cutoff. Thus zero energy is not an allowed state,
and the Fermi excitations are all at the left edge. 
\item[(ii).] $\frac{1}{2}-\frac{1}{\lambda} < \xi < 1- \frac{1}{\lambda}$. 
The filled interval is again the same as in (i) but the momenta pass through zero and thus the density of states is non-zero at zero energy. The excitations will again be at the left edge. 
\item[(iii).]  $1- \frac{1}{\lambda} <\xi < 0$.
The filled interval is now indeed $[-\frac{1}{2}, \frac{1}{2}]$. Now we can have excitations at both left and right Fermi edges. 
\item[(iv).] $0 < \xi  < \frac{1}{2}$. The filled interval is now $[\xi - \frac{1}{2}, \xi + \frac{1}{2}]$. 
The Fermi edge is now only on the right and we have excitations only there. We still have zero energy as an allowed state. 
\item[(v).]
$\frac{1}{2} < \xi $. The filled interval  is the same as in (iv), but the momenta are now all on the right and zero energy is no longer allowed. The Fermi edge continues to be on the right as in (iv). 
\end{enumerate}

Actually, the analysis in regions (iv) and (v) will not be much different and those in regions 
(i) and (ii) can be related to this by a simple change of variables. The region (iii) requires separate treatment. Hence let us first assume that $\xi$ is in regions (iv) or (v) i.e., $\xi>0$ and see how to develop the low temperature expansion there.  

To carry out the analysis we will adapt the Sommerfeld expansion in the theory of metals (see \cite{Ashcroft:1976uq} Chapter 2 and Appendix C, for instance). For any function $H(\tau)$ we will make an expansion 
\begin{equation}
\int_{\tau_L}^{\tau_R}H(\tau) \,\rho(\tau) \,d\tau = \int_{\tau_L}^{\tau_*}H(\tau) \,d\tau + \frac{c_1}{\hat \beta}\, H^{(1)}(\tau_*)+  \frac{c_2}{\hat \beta^2}\, H^{(2)}(\tau_*) +\cdots \ ,
\label{sommod}
\end{equation}	
with $H^{(i)}$ being determined in terms of $H$ and its derivatives. 
Here $\rho(\tau)$ is the Fermi-Dirac number density $$\rho(\tau) = \frac{1}{1+ e^{\frac{1}{2}\,{\hat \beta}\, (\tau^2 -\tau_*^2)}}$$ appearing in \eqref{zdet}. Introduce $K(\tau) = \int_{\tau_L}^{\tau} \, H(\tau')\, d\tau'$ i.e., 
$H(\tau) = \frac{d K}{d\tau}$, in terms of which we can write the l.h.s.\ of \eqref{sommod} as
\begin{equation}
\int_{\tau_L}^{\tau_R}\, H(\tau)\, \rho(\tau)\, d\tau = K(\tau_R) \,\rho(\tau_R)  +
\int_{\tau_L}^{\tau_R}\, K(\tau) \left(-\frac{d\rho(\tau)}{d \tau}\right) d\tau \ ,
\label{sommod1}
\end{equation}	
where we used $K(\tau_L) =0$ by construction. The analysis simplifies upon noting that as $\hat{\beta} \rightarrow \infty$ we have $\frac{d\rho(\tau)}{d \tau} \rightarrow 0$ outside a narrow interval in the vicinity of $\tau \approx \tau_*$. Thus we can expand $K(\tau)$ in a Taylor series in $(\tau -\tau_*)$ as 
\be
K(\tau) =K(\tau_*)+ (\tau- \tau_*)\, K^{'}(\tau_*) +\frac{1}{2} (\tau- \tau_*)^2\, K^{''}(\tau_*) +\cdots \ .  
\ee

Let us focus on the last term on the r.h.s.\ of \eqref{sommod1} and employ the Taylor expansion for $K(\tau)$. 
We introduce the modified variable  $v= \frac{1}{2}\,{\hat \beta}\, (\tau^2 -\tau_*^2)$. Then, for $\frac{v}{\hat{\beta}} \ll 1$,
we have from the definition of $v$
\begin{equation}
\tau- \tau_* = \frac{v}{\hat{\beta} \,\tau_*}\left(1 + \frac{\tau- \tau_*}{2\,\tau_*}\right)^{-1} \approx 
\frac{v}{\hat{\beta} \,\tau_*}\left(1 -\frac{v}{2\,\hat{\beta}\, \tau_*^2} +\cdots\right) \ .
\label{ytauexp}
\end{equation}	
Therefore using the fact that 
\be
-\frac{ d\rho}{d \tau}\, d \tau = \frac{e^{v}}{(1+ e^{v})^2}\,  dv
\ee
we can write the terms in the Taylor expansion as 
\begin{align}
&\int_{\tau_L}^{\tau_R}K(\tau) \left(-\frac{d\rho(\tau)}{d \tau}\right) d\tau  \approx 
K(\tau_*)\left[\rho(\tau_L) -\rho(\tau_R)\right]
\nonumber \\
& \qquad\quad  +\; \frac{K'(\tau_*)}{\hat{\beta} \,\tau_*} \int_{v_{\rm min}}^{v_{\rm max}}\, v \left(1 -\frac{v}{2\,\hat{\beta} \,\tau_*^2}\right)\frac{e^{v}}{(1+ e^{v})^2}\, dv 
+\frac{K''(\tau_*)}{2\,\hat{\beta}^2\, \tau_*^2} \int_{v_{\rm min}}^{v_{\rm max}}\, v^2\frac{e^{v}}{(1+ e^{v})^2} \,dv\,. 
\label{taylexp}
\end{align}	
Here $v_{\rm min} \equiv \frac{1}{2}\,{\hat \beta}\, (\tau_L^2 -\tau_*^2)$ and $v_{\rm max} \equiv
\frac{1}{2}\,{\hat \beta}\, (\tau_R^2 -\tau_*^2)$. For {\it generic} $\lambda$ (i.e., not close to $\lambda=1$) 
and for low enough 
temperatures, we have $\tau_R < \tau_*$ since $\tau_* \approx \tau_F < \tau_R$ (as we will self-consistently verify),
as well as for  $|\tau_L| < \tau_*$. Thus up to exponentially small corrections we can replace 
$v_{\rm min} \rightarrow -\infty$ and $v_{\rm max} \rightarrow +\infty$.  These conclusions will be altered when we take $\lambda \rightarrow 1$; we will delineate the changes in that limit later.

The integrals in \eqref{taylexp} can then be easily performed. Since $\frac{e^{v}}{(1+ e^{v})^2}$ is an even function, the odd powers of $v$ do not contribute in the integral and the only surviving terms to this order are
\begin{equation}
\int_{\tau_L}^{\tau_R}K(\tau) \left(-\frac{d\rho(\tau)}{d \tau}\right) d\tau  \approx K(\tau_*) 
+\frac{\pi^2}{3} \; \frac{\tau_*\,K''(\tau_*)-K'(\tau_*)}{2\, \hat{\beta}^2\, \tau_*^3} \ .
\label{taylexp2}
\end{equation}	
Here we have used the fact that $\rho(\tau_L) =1$ and $\rho(\tau_R) =0$ up to exponentially small corrections in $\hat{\beta}$. 

We apply this general formula in the case where $H(\tau)=1$, i.e., $K(\tau)= (\tau -\tau_L)$. 
Combining \eqref{sommod1} and \eqref{taylexp2} we have 
\begin{equation}
1= \int_{\tau_L}^{\tau_R}\, \rho(\tau) d\tau = (\tau_*-\tau_L) -\frac{\pi^2}{6} \,\frac{1}{\hat{\beta}^2\,\tau_*^3}\ ,
\label{numbexp}
\end{equation}	
leading to 
\begin{equation}
\tau_* \approx  \tau_L+1 +\frac{\pi^2}{6} \,\frac{1}{\hat{\beta}^2\,\tau_*^3}\approx \tau_F
+\frac{\pi^2}{6} \,\frac{1}{\hat{\beta}^2\,\tau_F^3} \ .
\label{cpotexp}
\end{equation}	
Thus we find that the leading temperature correction shifts the chemical potential {\it upwards} by a piece proportional to $T^2$. 
As we stressed earlier, this is the generic $\lambda$ low temperature behavior. We can also systematically find the 
low temperature expression for the free energy given in \eqref{Qpot2} in a similar way. Now $H(\tau) \propto \tau^2$ and 
therefore $K(\tau) \propto (\tau^3-\tau_L^3)$, and hence the $K^{''}(\tau_*)$ term in \eqref{taylexp} will also contribute. 
However the leading correction will still be proportional to $T^2$; explicitly one finds
\begin{equation}
 F_N(\lambda, \xi, {\hat \beta}) = N\left(-\frac{1}{24} -\frac{1}{2} \, \xi^2 + \frac{\pi^2}{6\,\hat{\beta}^2}\;\frac{1}{\tau_F} \right)\ .
\label{}
\end{equation}	
Since we have $\tau_F = \xi+ \frac{1}{2}$ in the domain $\xi > 0$ we learn that the free energy is independent of $\lambda$ at low temperatures. Note that in deriving the above we have assumed $T \ll \{\lambda, 1-\lambda\}$ making this analysis complementary to the discussion in \S{\ref{sec:5.2}.

To obtain the canonical free energy we still need to integrate over the auxiliary variable $\xi$. Before doing so however we should examine the other regions in $\xi$-space.  Firstly, we note that the behaviour for $\xi < \frac{1}{\lambda} -1 $, i.e., in the intervals (i) and (ii) can be obtained by a simple trick. Consider for definiteness the interval (i): we have at zero temperature filled momenta in the interval between  $\tau_F=\tau_R-1 = \xi - \frac{1}{2} +\frac{1}{\lambda}=
\xi +\frac{1}{2} +\frac{1-\lambda}{\lambda}$. The lower limit on momenta is now $\tau_L= \xi - \frac{1}{2} < \tau_F$ and so the excitations are on the left edge. However by a change of variables we can bring this to a more familiar form. Relabeling $\tau' = -\tau$ and $\tau_{L,R}' = -\tau_{R,L}$ etc., together with $\xi' = -\xi -\frac{1-\lambda}{\lambda}$ we reduce the problem to exactly what we have considered above, except in terms of the primed variables. In particular, we simply need to write the  answers in terms of the primed variables and express those in terms of $\xi$ using the above dictionary to get the final values. Similar considerations apply to region (ii). Finally, in the interval (iii), the Fermi points are at $\pm \frac{1}{2}$ at zero temperature. This means that we can adapt the textbook story of the Sommerfeld expansion quite straightforwardly (relative to $\tau = \pm \frac{1}{2}$ we have $\tau_{L,R}\to \mp \infty$ in our regime of operation). 

Putting these results together we note that the Fermi level is always given by \eqref{cpotexp}, with the value of $\tau_F$ appropriate for the interval of $\xi$ under consideration. The free energy as a function of $\xi$ is then
\begin{equation}
\frac{1}{N}\, F_N(\lambda, \xi, {\hat \beta}) =
\begin{cases}
& -\frac{1}{24} -\frac{1}{2} \, \xi^2 + \frac{\pi^2}{6\,\hat{\beta}^2}\;\frac{1}{\xi+\frac{1}{2}} \,,\hspace{3.2cm}  \xi > 0 \\
&  -\frac{1}{24} \,,\hspace{5.5cm} \frac{\lambda-1}{\lambda} <\xi <0\\
&-\frac{1}{24} -\frac{1}{2} \, \left(\xi +\frac{1-\lambda}{\lambda}\right)^2 + \frac{\pi^2}{6\,\hat{\beta}^2}\;\frac{1}{\xi-\frac{3}{2}+\frac{1}{\lambda}}\,, \hspace{1.1cm} \xi < \frac{\lambda-1}{\lambda}\ .
\end{cases}
\label{}
\end{equation}	
It is clear from these expressions that the correction to the zero temperature free energy $F(\lambda, \xi)|_{T=0} = -\frac{N}{24}$ is extremely benign. Furthermore, the saddle point evaluation of the $\xi$ integral to obtain the canonical free energy of the system can be argued to be dominated by this zero temperature contribution. In the end we simply obtain
\begin{equation}
F_N(\lambda,\beta) = \frac{\lambda^2}{24\, N}
\label{finalF}
\end{equation}	
upon completing the saddle point integration with $\xi_\text{saddle} =0$ (and accounting for the fluctuation determinant). The factor of $\lambda^2$ can be traced back to our rescaling of variables.  
 
While the analysis described above is valid for small $T$ it fails when ${\hat \beta}\, (1-\lambda) \simeq 1$. In this limit we have the potential for the Fermi surface to interfere with the upper end of the momentum integration. This follows from the observation that $\tau_R -\tau_L \sim 1+ (1-\lambda)$ in this limit. 
 To ascertain whether this leads to a non-trivial result for the free energy, we have also examined the Sommerfeld expansion in the double scaling limit ${\hat \beta} (1-\lambda)$ fixed with ${\hat \beta} \gg 1 $ and $(1-\lambda) \to 0$. At first sight the modifications in the Sommerfeld expansion indicate that we could get a correction that scales linearly in $T$ as opposed to the quadratic $T^2$ corrections we encountered above. Indeed, this is the case for the Fermi energy: we find that $\tau_*-\tau_F = T \, g({\hat \beta}, 1-\lambda)$ in this range of parameters. For example when $\xi > 0$ or $\xi < \frac{\lambda-1}{\lambda}$ we find that $g({\hat \beta}, 1-\lambda) = \frac{1}{\tau_F}\, \log (1-e^{-{\hat \beta}\,(1-\lambda)\,\tau_F})$. In the intermediate region (iii) there is a rather complicated expression for the Fermi level.  However, when it comes to the free energy the various contributions at linear order in $T$ cancel out in a non-trivial fashion. In fact, the end result is quite boring: we find $F_N(\lambda\, \xi, {\hat \beta}) =-\frac{N}{24} - \frac{N}{2}\,\xi^2+ {\cal O}(T^2)$ despite non-trivial intermediate results for $\xi \in {\mathbb R}$. From here of course we simply end up obtaining \eqref{finalF}, which has no features. We should note that we have not carried out an explicit check at ${\cal O}(T^2)$; it is not impossible that there is some interesting effect lurking in this region.



\providecommand{\href}[2]{#2}\begingroup\raggedright\endgroup

\end{document}